# Understanding Heat Transport Mechanisms in Optically Transparent Thermal Loss Mitigators

Domala Sai Suhas[a#] and Vikrant Khullar[a,*]
[a]Mechanical Engineering Department, Thapar Institute of Engineering & Technology, Patiala-147004, Punjab, India
*Corresponding author. Email address: vikrant.khullar@thapar.edu
#Present address: Department of Mechanical and Aerospace Engineering, University of Tennessee, Knoxville, TN 37996, U.S.A.

**ABSTRACT**
Optically transparent thermal loss mitigators have recently seen renewed research interests owing to their increasing relevance in the realms ranging from smart windows, efficient greenhouse designs and high-performance-low-cost solar thermal systems. In depth understanding of the heat transport mechanisms and their quantification is crucial for building efficient opto-thermal management strategies for optimization of the aforementioned systems. The present work serves to identify and quantify the key heat transfer mechanisms operative in a host of optically transparent thermal loss mitigators. In particular, comprehensive experimental modelling frameworks have been developed to investigate the efficacy of carbon dioxide gas ($CO_2$), air, vacuum (0.07mbar), transparent heat mirrors (Indium tin oxide coated glass) and aerogels (silica-based) in mitigating thermal losses. Detailed and careful experimental modelling reveals that it is imperative to employ more than one thermal loss mitigator and choose correct absorber surface orientation (relative to the irradiation direction) to maximize thermal loss mitigation. Magnitude of absorber surface stagnation temperature has been employed as the figure of merit to quantitatively compare various optically transparent thermal loss mitigators. Under un-evacuated conditions, $CO_2$ has emerged as potent alternative to more sophisticated optically transparent thermal loss mitigators like aerogels and transparent heat mirrors. Enhancements (relative to air) on the order of 2%-7%, 46%-84%, 57%-84% and 66%-86% are observed in case of $CO_2$, vacuum, transparent heat mirrors (vacuum) and aerogel (vacuum) respectively.

## 1. INTRODUCTION

Optically transparent thermal loss mitigators (OTTLMs) are particularly important in applications, wherein, the objective is to allow a region of the electromagnetic spectra (normally Vis/Vis-NIR) to pass through but at the same time block the MIR region of the electromagnetic spectra. Furthermore, it should have attributes such as low thermal conductivity and/or should inhibit bulk fluid motion (to contain convection). Among its host of potential application areas, the realm of solar thermal systems in particular holds promise.

The key strategy for harnessing solar thermal energy in surface-absorption based solar-thermal systems involve efficient absorption of solar energy, and at the same time, mitigation of thermal losses from the absorbing surface [1]. Figure 1 details characteristics of ideal absorbing surface as well as ideal OTTLMs. Incumbent solar thermal technologies (with the aim to realize efficient photo-thermal energy conversion) essentially employ solar selective surface as the absorbing surface (for efficient absorption of solar energy and at the same time curbs radiative losses) in conjunction with OTTLMs such as vacuum technology (for curbing conduction and convection losses). Although, a proven technology, it suffers from significant initial capital investment and long-term operational and maintenance issues. Therefore, there have been numerous attempts to explore alternative routes to harness solar thermal energy. In this direction, recently, a few researchers have reported that transparent insulating materials (TIMs) such as transparent aerogel [2-6] could be employed, wherein, it allows the sunlight to reach the absorbing surface but mitigates thermal losses, hence resulting in efficient photo-thermal energy conversion. For instance, absorber surface temperatures greater than 200 °C under unconcentrated sunlight has been reported by Zhao et al. [4]. Berquist et al. [5] has further increased the attainable absorber temperatures to 700°C under low solar concentration ratio ($C = 10$) by seeding Indium Tin oxide (ITO) nanoparticles into the aerogel and has recently demonstrated scalability [6] as well.

Transparent heat mirrors (THMs) are another class of materials which have been explored as OTTLMs in solar thermal systems. Typically, THMs are employed as cover materials for the absorbing surface in order to curb radiative losses (in particular). Herein, the sunlight is able to pass through the THM, however, the infrared radiations emanating from the absorber surface are reflected back to the absorber, hence reducing the radiative losses [7-10].

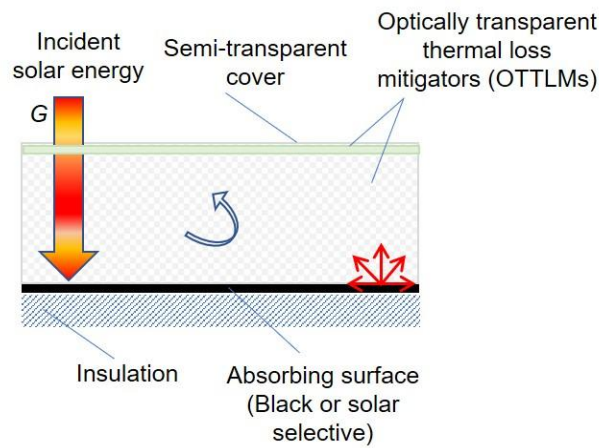

| Component | Material | Ideal spectral radiative characteristics | Ideal thermal conductivity $(k)\ Wm^{-1}K^{-1}$ | Ideal convection heat transfer coefficient$(h_{conv})$ $Wm^{-2}K^{-1}$ |
|---|---|---|---|---|
| Absorbing surface | Black surface | $\alpha_\lambda = \varepsilon_\lambda = \begin{cases} 1, \lambda < \lambda_{cut\ off} \\ 1, \lambda \geq \lambda_{cut\ off} \end{cases}$ | $\infty$ | Not applicable |
| | Solar selective surface | $\alpha_\lambda = \varepsilon_\lambda = \begin{cases} 1, \lambda < \lambda_{cut\ off} \\ 0, \lambda \geq \lambda_{cut\ off} \end{cases}$ | | |
| OTTLMs | OTTLMs | $\tau_\lambda = \begin{cases} 1, \lambda < \lambda_{cut\ off} \\ 0, \lambda \geq \lambda_{cut\ off} \end{cases}$ | 0 | 0 |

(b)

Fig. 1 (a) Schematic diagram of a typical solar thermal system employing OTTLMs, and (b) ideal opto-thermal characteristics of absorbing surface and OTTLMs.

Although, aerogels and THMs have emerged as potent candidates for harnessing solar thermal energy; but these are still in their nascent stages of development (see Fig. 2). Therefore, further investigation into the operational heat transfer mechanisms in the aforementioned OTTLMs is warranted to optimize their effectiveness and realize their deployment in real world systems. The present work is a significant step in this direction, wherein, we have developed mechanistic experimental modelling frameworks to identify, quantify and engineer the energy transport mechanisms in the aforementioned OTTLMs. Additionally, for the first time (to the best of our knowledge) the candidature of carbon dioxide ($CO_2$) as a potential OTTLM has been investigated. $CO_2$ being transparent in the visible wavelength band and absorbing in certain IR wavelength bands (characteristics of a typical green-house gas) can potentially trap the IR radiations emitted from the absorber surface, hence mitigating radiative losses in particular. Instructively, the direction of irradiation, material and the location of the absorber surface has also been carefully engineered to control the heat transport mechanisms.

| SURFACE ABSORPTION BASED SOLAR THERMAL SYSTEMS | | OTTLM | | | | |
|---|---|---|---|---|---|---|
| | | $CO_2$ | Transparent heat mirror (THM) | | Aerogel | |
| | | | @ air | @ vacuum | @ air | @ vacuum |
| RECEIVER CONFIGURATION | Incident solar energy ↓ OTTLM / Insulation / Solar selective surface | | Fan and Bachner, 1976 [T] | Khullar et al., 2018 [T] | | |
| | Incident solar energy ↓ OTTLM / Insulation / Black surface | | | | Berquist et. al., 2025 [T, E]; Berquist et. al., 2020 [T, E]; Zhao et. al., 2019 [T, E]; Gunay et. al., 2018 [T]; McEnaney et. al., 2017 [T] | |
| | Insulation / Solar selective surface / OTTLM ↑ Incident solar energy | | SCOPE OF THE PRESENT WORK | | | |
| | Insulation / Black surface / OTTLM ↑ Incident solar energy | | | | | |

Fig. 2 Selected reported works pertinent to OTTMs and scope of the present work.

## 1.1   Thermal loss transport mechanisms in OTTLMs

As noted earlier, OTTLMs need to be transparent in the desired wavelength band and at the same time need to curb radiation, convection and conduction losses. Figure 3(a) details the attributes of various OTTLMs in this regard. Clearly, no single OTTLMs has all the desired characteristics and necessitates trade-off/engineering to realize the required opto-thermal characteristics. To engineer such diverse opto-thermal characteristics, it is imperative to carefully employ more than one OTTLMs such that they complement and or supplement each other (see Fig. 3(b)). Additionally, relative position of the absorbing material and direction of irradiation also need to be carefully chosen to control convection losses in particular.

| Thermal loss transport mechanism | | Radiation | | Conduction | Convection | |
|---|---|---|---|---|---|---|
| | | Short wavelength radiations | Long wavelength radiations | | | |
| Impacting parameter | | Transmissivity | Transmissivity | Thermal conductivity | Direction of irradiation | |
| | | | | | Irradiation from bottom | Irradiation from top |
| OTTLMs | Vacuum | Transparent | Transparent | Very low | Negligible | |
| | Air | Transparent | Transparent | | | |
| | $CO_2$ | Transparent | Absorbing (in certain IR wavelength bands) | | Negligible | Significant |
| | Aerogels | Transparent (depends on particle size) | Absorbing (in most of the IR wavelength region) | | Not applicable | |
| | Heat mirrors | Transparent | Reflective (in most of the IR wavelength region) | | Not applicable | |

(a)

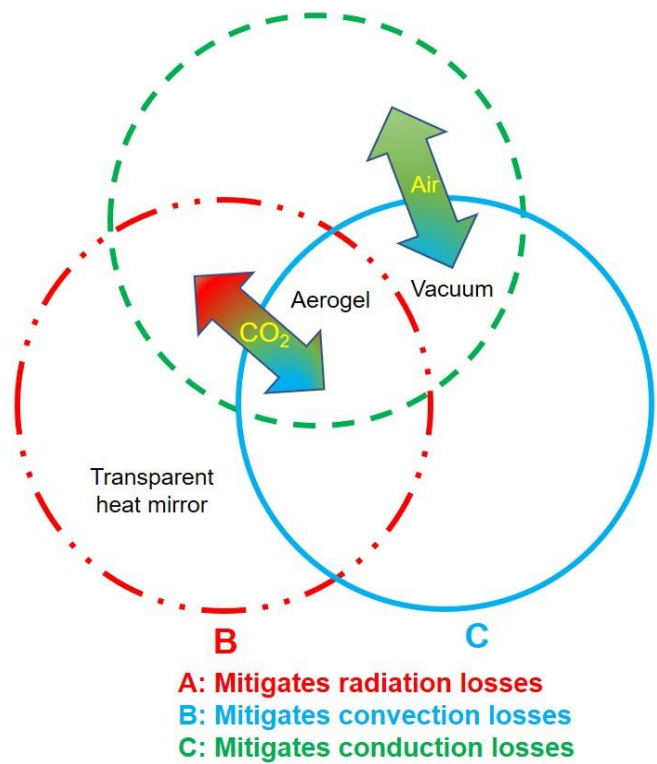

(b)

Fig. 3(a) Thermal loss transport mechanisms relevant o various OTTLMs, and (b) effectiveness of various OTTLMs in mitigating radiation, convection and conduction losses.

## 2. OPTICAL PROPERTIES OF ABSORBING MATERIALS AND OTTLMs

In the present work, to develop experimental modelling frameworks for quantifying the thermal loss mechanisms, various absorbing surfaces and OTTLMs have been employed. This section aim to quantify and analyse the optical characteristics of absorbing surfaces and OTTTLMs.

### 2.1 Optical properties of spectrally black and solar selective surfaces.

To single out and quantify radiative losses in the experiments, two different absorbing surfaces have been employed. The two surfaces being "spectrally black" (Spectral black$^{TM}$ foil, Acktar Advanced Coatings) and "solar selective" (Black Chrome solar selective coated copper substrate, Solchrome Private Ltd.) respectively. These have nearly similar optical characteristics in the UV-Vis-NIR region (see spectral values in Fig. 4(a) and solar weighted absorptivity ($\alpha_{sw}$) values defined by Eq. (1) in Fig. 4(c)) but distinctly different optical properties in the MIR region (see Fig. 4(a)). Thus, allowing control over effective emissivity ($\epsilon_{eff}$) (see Fig. 4(b)) and hence the magnitude of the radiatively losses from the absorbing surface.

$$\alpha_{sw} = \frac{\sum_{\lambda=0.3\mu m}^{\lambda=4\mu m} S_\lambda \alpha_\lambda}{\sum_{\lambda=0.3\mu m}^{\lambda=4\mu m} S_\lambda} \tag{1}$$

$$\epsilon_{eff} = \frac{\sum_{\lambda=0.19\mu m}^{\lambda=14\mu m} E_{b,\lambda,T} \alpha_\lambda}{\sum_{\lambda=0.19\mu m}^{\lambda=14\mu m} E_{b,\lambda,T}} \tag{2}$$

Where, $S_\lambda$ and $\alpha_\lambda (= \epsilon_\lambda)$ are the spectral solar irradiation (AM 1.5) and absorptivity values; $E_{b,\lambda,T}$ being the spectral emissive power (computed using Planck's distribution law).

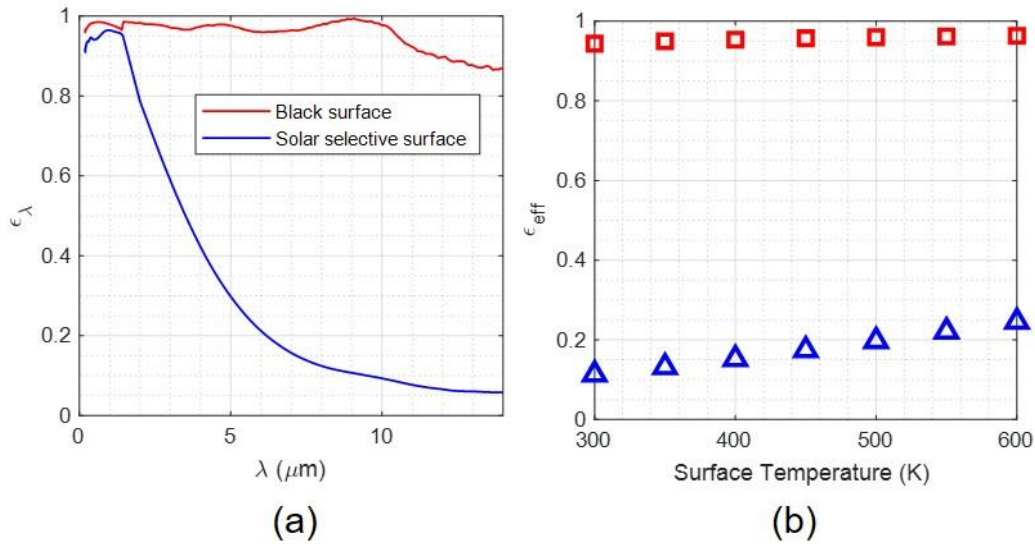

Fig. 4(a) Spectral emissivity ($\epsilon_\lambda$), (b) effective emissivity, and (c) solar weighted absorptivity of black and solar selective surfaces. For black surface spectral reflectance measurements: 0.190$\mu$m to 1.4$\mu$m (UV–Vis–NIR spectrophotometer with an integrating sphere (Shimadzu, UV-2600i) has been employed); 1.45$\mu$m to 14$\mu$m data taken from Ref. [11]. For solar selective surface spectral reflectance measurements: 0.190$\mu$m to 1.4$\mu$m (UV–Vis–NIR spectrophotometer with an integrating sphere (Shimadzu, UV-2600i) has been employed); 2$\mu$m to 20$\mu$m (FTIR spectrophotometer with integrating sphere (PerkinElmer, Frontier FTIR) has been employed.

## 2.2 Optical properties of aerogel, THMs, and $CO_2$

Among the OTTLMs listed in Fig. 3(a), air and vacuum are nearly transparent in the UV-Vis-NIR-MIR wavelength region and therefore have not been investigated for their optical properties. However, the remaining OTTLMs namely aerogel, THMs, and $CO_2$ have distinct spectral optical characteristics and have therefore been investigated using empirical correlations, UV-Vis spectrometer (Sekonic, C-800 Spectro Master), UV-Vis-NIR (Agilent, Carry 5000 and Shimadzu, UV-2600i) and FTIR (PerkinElmer, Frontier FTIR) spectrophotometers.

### *2.2.1 Optical properties in UV-Vis-NIR wavelength region.*

***Total spectral transmittance measurements in the visible region (0.38µm to 0.78µm)***
Given the fact that silica aerogel (Hydrophobic silica aerogel, Aerogel Technologies®) is very fragile, it was difficult to hold it in the sample holder of a regular spectrophotometer equipped with integrating sphere. Therefore, an in-house experimental setup was developed to measure the total spectral transmittance in the visible region of the electromagnetic spectra. For this purpose, a solar simulator (SciSun-300, Solar Simulator, Class AAA, 300W, with AM1.5 air mass filter) was employed as the light source and a spectrometer (Sekonic, C-800 Spectro Master) was employed as the detector. The spectrometer was inserted into a cardboard box with a square hole cut to form the aperture (see Fig. 5(a)). The sample to be tested was then placed over it so that the radiations from the source pass through the sample before reaching the detector (see Figs. 5(b), 5(c), 5(d) and 5(e)). Furthermore, to compute the transmittance of the samples, and to take a baseline measurement, measurements were also made without keeping any sample over the aperture.
Figure 5(f) show the spectral transmittance curves for aerogel (with and without reflector) and THM (ITO coated glass substrate (surface resistivity 70-100 $\Omega$/sq., Merck), radiation incident on coated/uncoated sides). Clearly, silica aerogel has much lower transmittance compared to the ITO coated glass owing to scattering of the incident light by the silica particles.

***Direct spectral transmittance measurements in the UV-Vis-NIR region (0.25µm to 2.5µm)***
To quantify the scattering magnitude, additional measurements of direct transmittance were made using universal measurement accessory (UMA) attached to UV-Vis-NIR spectrophotometer (Agilent, Carry 5000). Spectral direct transmittance measurement involves placing the sample in between the source and the detector such that only the un-scattered direct radiation reaches the detector (see Fig. 6(a)). Herein, no clamping was required to hold the sample, thus allowing measurement of spectral direct transmittance of fragile samples like aerogel as well. Figure 6(b) shows direct spectral transmittance curves (0.25$\mu$m to 2.5$\mu$m) for aerogel and ITO coated glass substrate; clearly, in the entire wavelength range aerogel has significantly lower transmittance. The effect of scattering becomes clearer if the total and direct transmittance curves are compared side by side (see Fig. 6(c)). In case of ITO coated glass substrate, the two set of curves almost coincide – indicating negligibly small scattering. On the other hand, for aerogel, these curves don't coincide – indicating significant scattering.

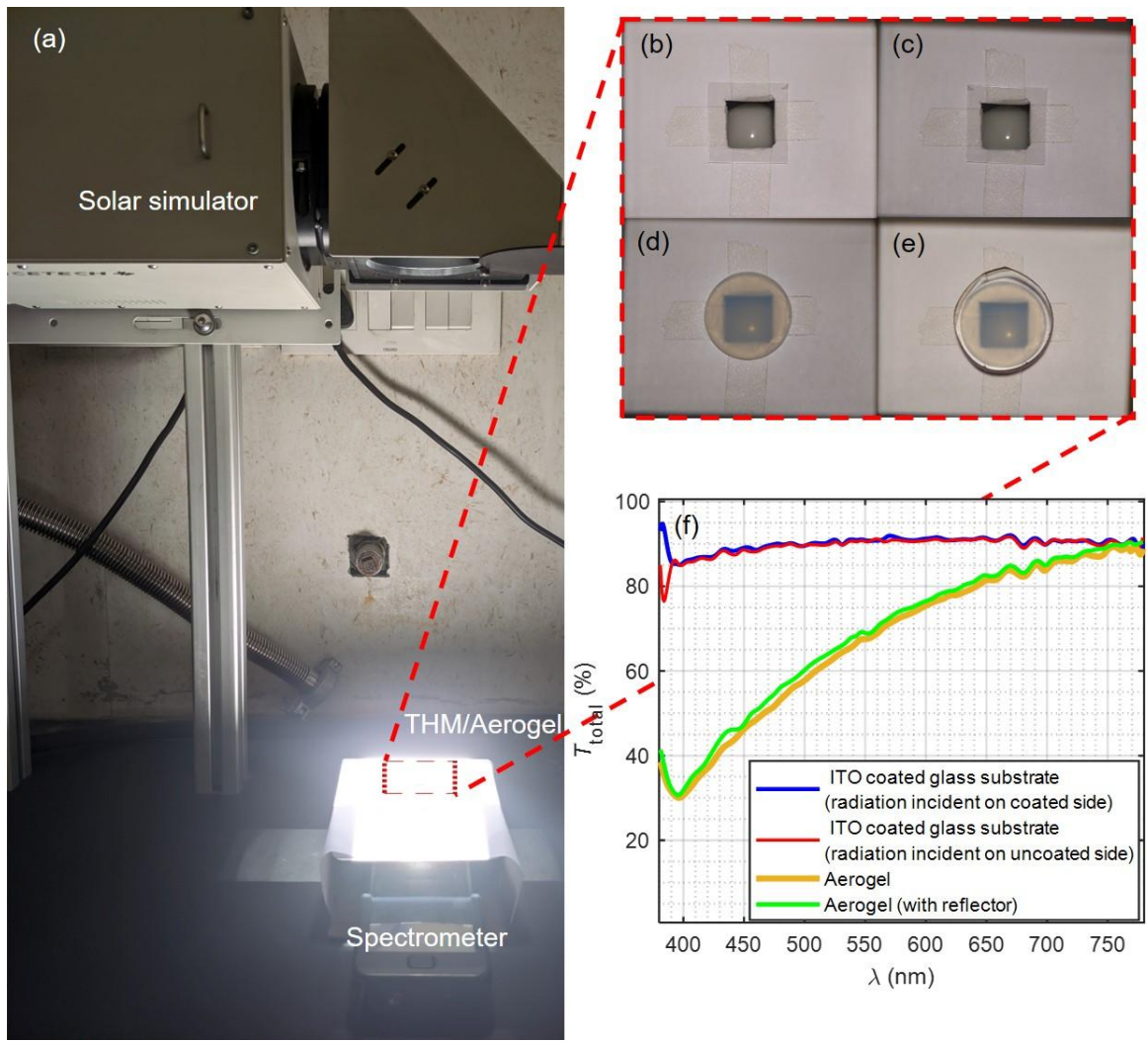

Fig. 5(a) Photograph of the experimental setup employed for measuring spectral optical properties of (b) ITO coated glass substrate (radiation incident on coated side), (c) ITO coated glass substrate (radiation incident on uncoated side), (d) silica aerogel, and (e) silica aerogel surrounded by reflecting surface (3M$^{TM}$ solar mirror film). Spectral total transmittance as measured by spectrometer.

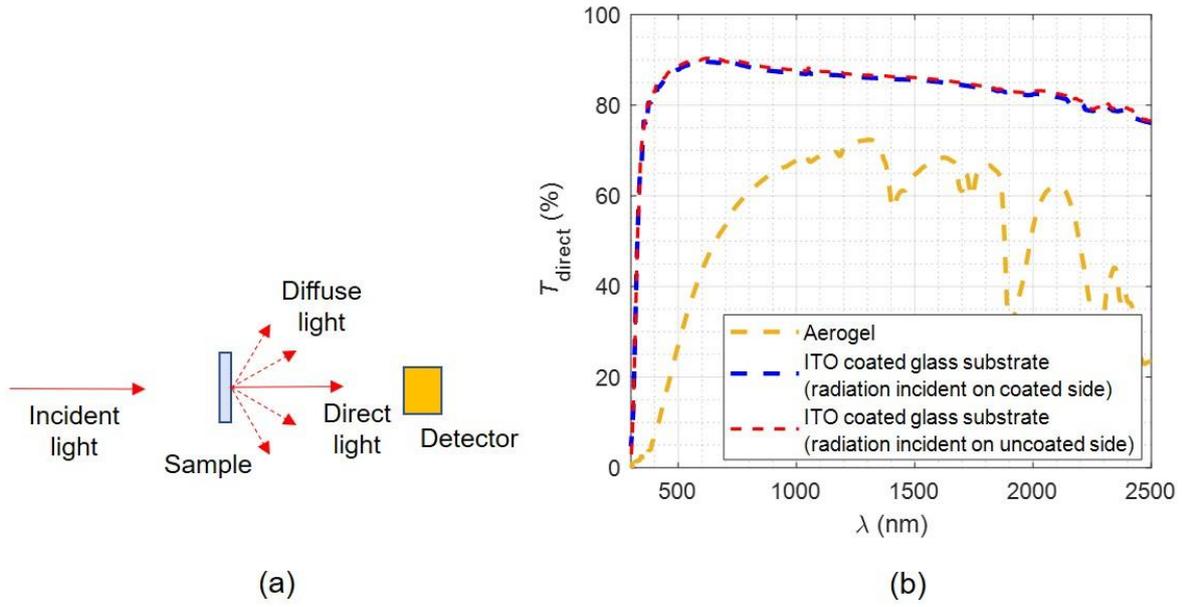

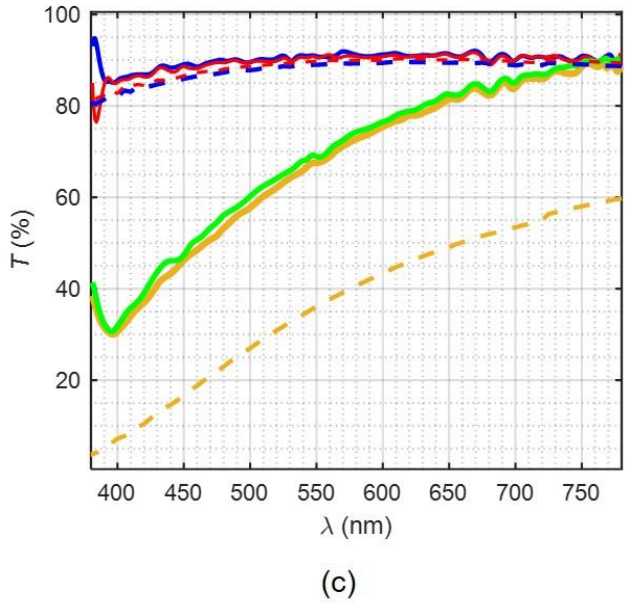

Fig. 6(a) Schematic diagram of the basic operating principle in measuring direct transmittance, (b) spectral direct transmittance curves from 0.2$\mu m$ to 2.5$\mu m$, and (c) comparison of direct and total spectral transmittance.

### *2.2.2 Optical properties in MIR wavelength region.*

### *Optical properties of $CO_2$*

The spectrally banded optical properties of carbon dioxide (which have been computed using correlations detailed in Table. 1) depend on the mass path length ($w$), given by Eq. 3 [12]; which in turn is a function of total pressure ($P$ = 0.9737 atm., in the present work), absolute temperature ($T$), mole fraction of the gas ($x$ =1, in the present work), physical path length ($L$ = 0.080m, in the present work), and the universal gas constant ($R$ = 8.314 Jmol$^{-1}$K$^{-1}$)

$$w = \frac{xPL}{RT} \tag{3}$$

For the two different temperature values, the mass path length values are given as $w_{@294K} = 142.1048$ gm$^{-2}$ and $w_{@555K} = 75.27714$ gm$^{-2}$. Furthermore, equivalent broadening pressure is given by Eq. 4 as

$$P_E = P(1 + x + 0.5x^2 P) = 2.5P = 2.43421 \text{ atm.} \tag{4}$$

Table 1. Various absorption bands, their strengths and corresponding absorption coefficients correlations and values pertinent to $CO_2$.

| Band | Band strength | $T$ [K] | Band absorption correlation [$cm^{-1}$] | Absorption coefficient [$cm^{-1}$] |
|---|---|---|---|---|
| 15$\mu$m | Medium | 294 | -15+51 $log_{10}w$+22 $log_{10}P_E$ | 103.2829 $\pm$ 5 |
| | | 555 | -36+82 $log_{10}w$+19 $log_{10}P_E$ | 126.3863 $\pm$ 5 |
| 10.4$\mu$m | Weak | 294 | 0.0008w | 0.113684 $\pm$ 2 |
| | | 555 | 0.018w | 1.354989 $\pm$ 2 |
| 9.4$\mu$m | Weak | 294 | 0.001w | 0.142105 $\pm$ 2 |
| | | 555 | 0.02w | 1.505543 $\pm$ 2 |
| 5.2$\mu$m | Weak | 294 | 0.0016w | 0.227368 $\pm$ 2 |
| 4.8$\mu$m | Weak | 294 | 0.0082w | 1.165259 $\pm$ 2 |
| 4.3$\mu$m | Strong | 294 | 75 $w^{0.11} P_E^{0.08}$ | 138.9175 $\pm$ 7 |
| | | 555 | 93 $w^{0.11} P_E^{0.05}$ | 156.3989 $\pm$ 7 |
| 2.7$\mu$m | Medium | 555 | 13.5 $w^{0.5} P_E^{0.06}$ | 123.5512 $\pm$ 22 |
| 2.0$\mu$m | Weak | 294 | 0.07w | 9.947337 $\pm$ 15 |
| | | 555 | 0.082w | 6.172726 $\pm$ 15 |

Figure 7 shows that $CO_2$ has absorption bands spread across the entire MIR and does not have absorption bands in the UV-Vis-NIR. Thus, $CO_2$ has favourable optical characteristics which lends it to be employed as an OTTLM.

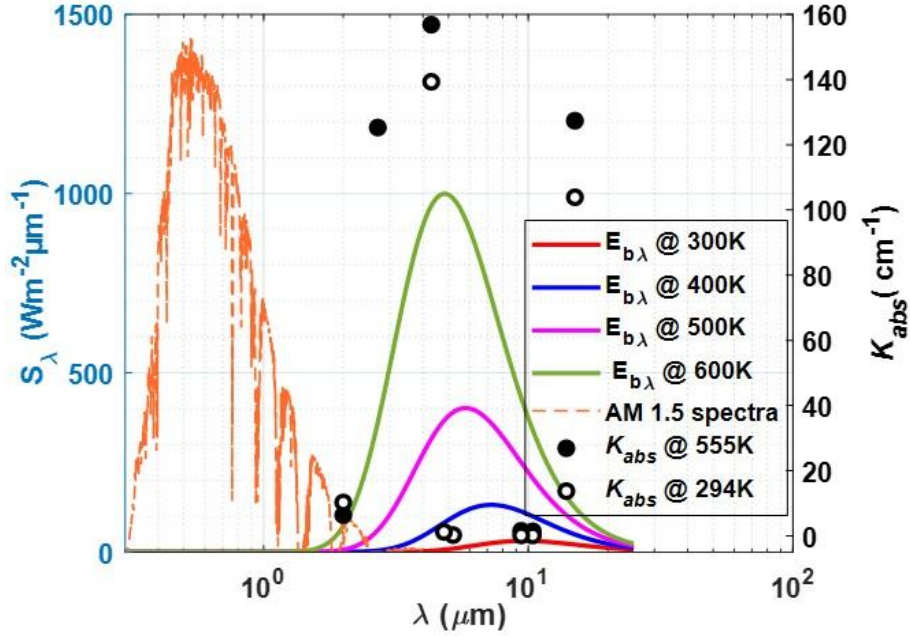

Fig. 7 Spectral incident solar radiation (AM 1.5), (b) spectral blackbody emissive power at various temperatures, and (c) spectral absorption coefficients of $CO_2$.

*Optical properties of ITO coated glass substrate*
Interestingly, optical properties of ITO coated glass substrate depend on whether the irradiation is incident on the coated or uncoated side. In other words, the coated side could either face the absorbing surface or could face the incident irradiation. Therefore, to understand and compare the optical characteristics of these THMs in the aforementioned two configurations, the optical properties were measured for two sides of THM. FTIR spectrophotometer with integrating sphere has been employed to measure spectral transmittance and reflectance. Although transmittance curves nearly coincide (see Fig. 8(a)), there is significant difference in the specular reflectance curves (see Fig. 8(b)). Thus, pointing out difference in absorptance (and hence emissivity values) of the coated and uncoated sides of the THM (see Fig. 8(c)). To quantify the effective emissivity values for a given absorber surface temperature, IR weighted emissivity has been computed using Eq. (5)

$$\epsilon_{IR,ITO} = \frac{\sum_{\lambda=2.5\mu m}^{\lambda=25\mu m} E_{b,\lambda,T} \alpha_\lambda}{\sum_{\lambda=2.5\mu m}^{\lambda=25\mu m} E_{b,\lambda,T}} \qquad (5)$$

Figure 8(d) shows IR weighted emissivity as a function of absorber surface temperature. For both the coated as well as uncoated sides, the IR weighted emissivity decreases with increase in absorber temperature. However, oppose to the coated side, in case of the uncoated side, the IR weighted emissivity is a strong function of absorber temperature. Clearly, the coated side of the THM (ITO coated substrate) has much reduced IR weighted emissivity (29% – 37% lower than the uncoated side); hence, effectively curbs radiative loss across the intermediate temperature range of the absorbing surface.

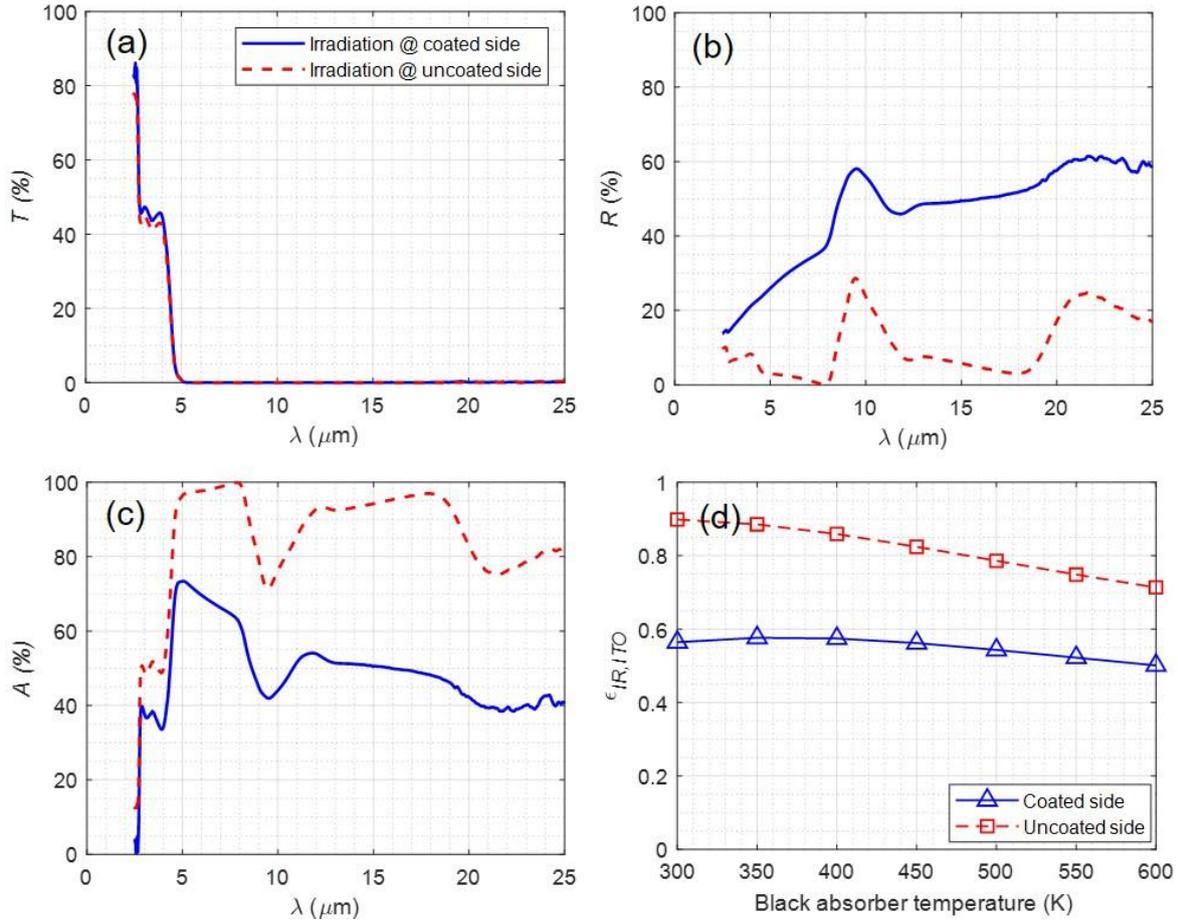

Fig. 8 (a) Optical properties of coated and uncoated sides of THM (ITO coated glass substrate) (a) measured spectral transmittance, (b) measured spectral reflectance, and (c) calculated spectral absorptance ($A = 1 - T - R$), and calculated IR weighted emissivity.

*Optical properties of aerogel*
Aerogel being very fragile (and due to the size constraints), it is not possible to hold it in the available sample holder for taking measurements employing integrating sphere. Therefore, only direct transmittance has been measured (that too by circumscribing aerogel with a reflecting cardboard ring (3M$^{TM}$ solar mirror film pasted on cardboard ring) to provide some rigidity to the aerogel sample, see Fig. 9(a)). Nevertheless, direct transmittance values do give us an idea of the wavelength window (in the MIR wavelength region) where the aerogel is transparent. Figure 9 shows the spectral direct transmittance of aerogel circumscribed by a reflective surface (reflecting ring). Clearly, aerogel is nearly absorbing in nearly the entire MIR of the electromagnetic spectra except for a transparent window viz., 3.5$\mu$m – 5$\mu$m (see Fig. 9(b)). Hence, aerogels are capable of curbing most of the radiation losses from the absorber surface.

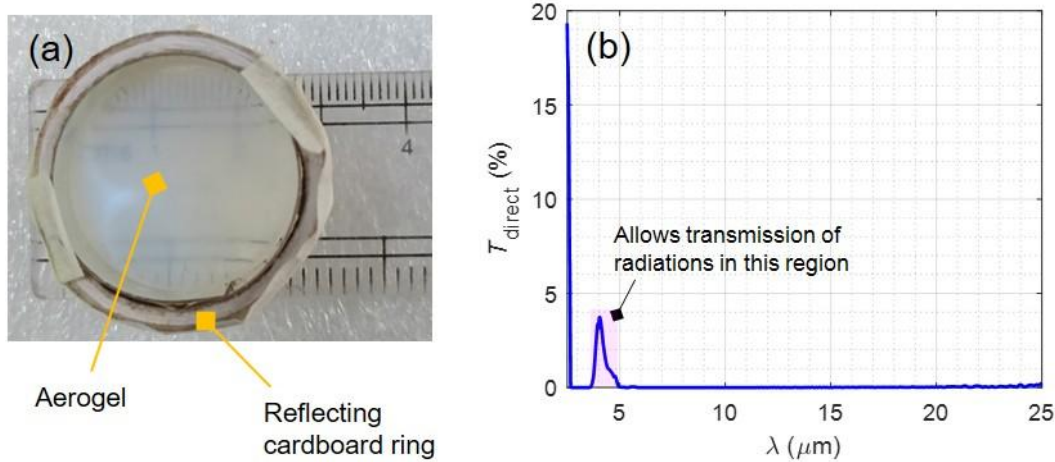

Fig. 9 (a) Aerogel sample circumscribed by a reflecting ring, and (b) spectral direct transmittance in MIR of the electromagnetic spectra.

## 3. EXPERIMENTAL MODELLING FRAMEWORK

Now that we have analysed the optical characteristics of the absorbing surfaces and OTTLMs, next, it is imperative to assess their thermal performance under optical heating. For the said purpose, a comprehensive experimental modelling framework has been developed to allow for thermal testing of various "absorbing surface-OTTLM" combinations.

### 3.1 Experimental setup: basic construction

A solar simulator (SciSun-300, Class AAA, 300W, Sciencetech Inc.) equipped with an air mass filter (AM1.5G-FT-3, Sciencetech Inc.) has been employed as the source for optical heating. An in-house stainless steel (SS304) chamber (see Fig. 10) has been fabricated to house various absorber surface-OTTLM combinations. The chamber has KF10 and KF25 flanges (each 02 in number) welded to its outer surface. One of the KF25 flanges is connected to KF25-KF40 adapter for the 26-pin feedthrough access for thermocouples ($K$-type). Another KF25 flange is used for creating vacuum and/or filling carbon dioxide ($CO_2$). One of the KF10 flanges is used for connecting the Pirani gauge and the other for connecting analogue pressure gauge. A stainless-steel plate (partition plate) was stuck to the chamber using epoxy-based clear adhesive (Bondtite, Astral Adhesives). The receiver is attached on to the top surface of this partition plate. Also, an insulated hollow cylindrical copper tube was inserted through the partition plate to allow for vacuum creation (and/or $CO_2$ gas filling) and to ensure equalization of pressure on the two sides of the plate (see Fig 10(a), 10(b), and 11(b)).

$K$-type thermocouples have been employed to measure spatial temperature distribution within the chamber, absorber surface temperature, glass cover temperature, ambient temperature and temperature on the other side of the partition plate. Furthermore, radiation shields (made of $3M^{TM}$ solar mirror film) have been employed to ensure that thermocouples' do not directly interact with the incoming radiations during optical heating (see Fig. 11).

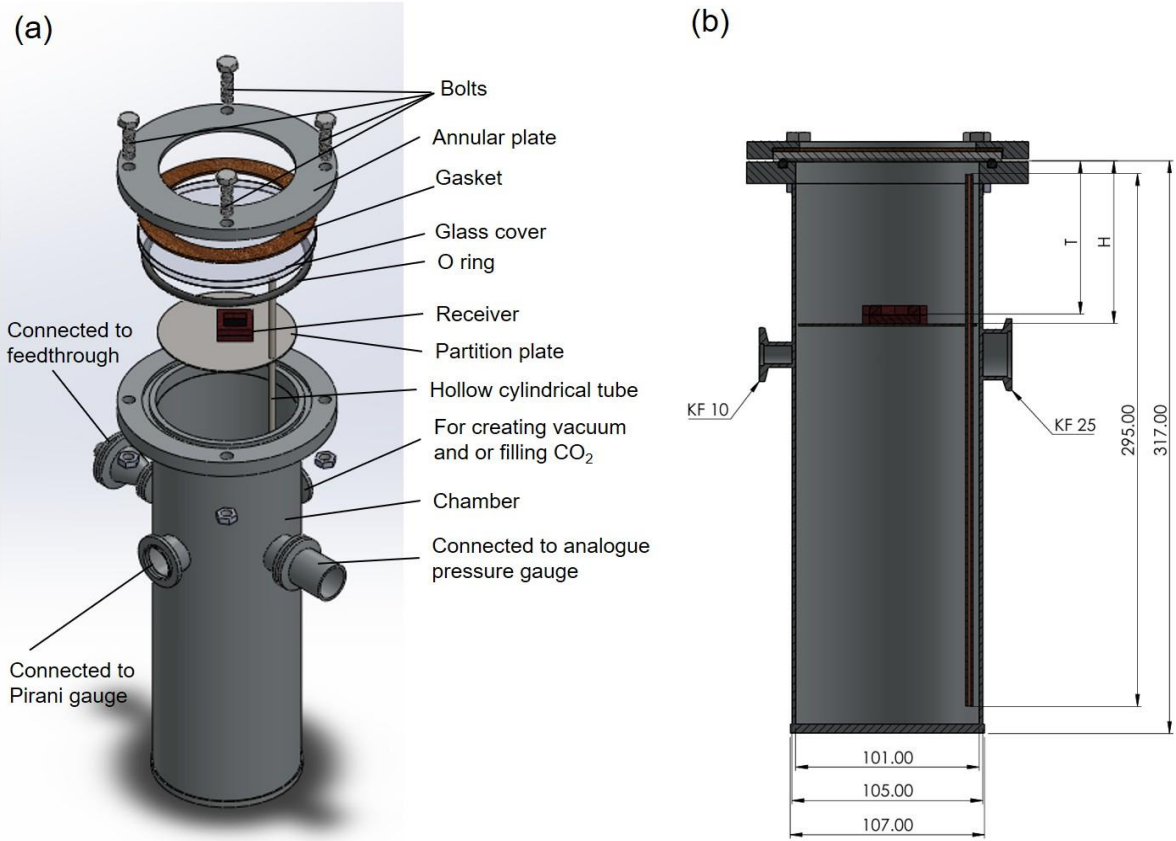
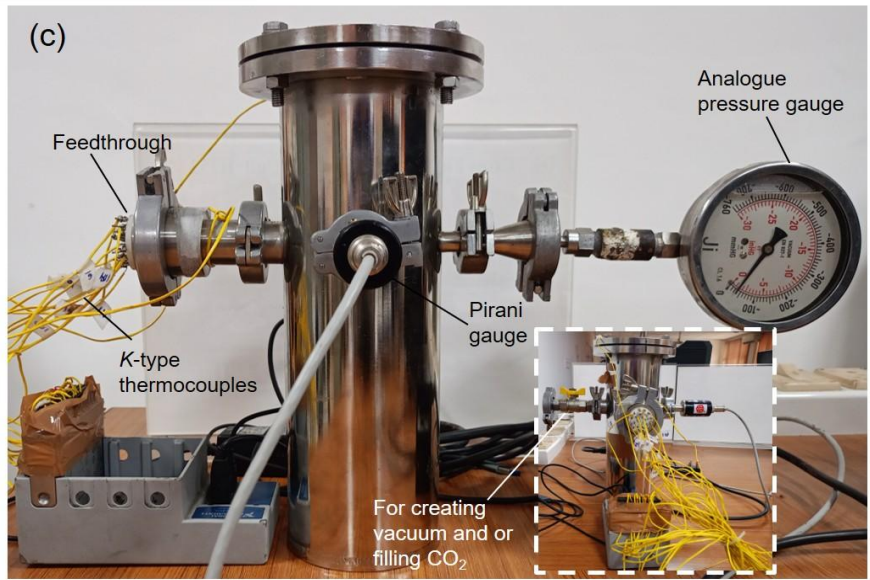

Fig. 10 Experimental setup chamber (a) Exploded view, (b) sectional view, and (c) picture of the of the chamber along with various attachments. The inset in Fig 10 (c) is the rotated view (90° anticlockwise, if viewed from top) of the chamber.

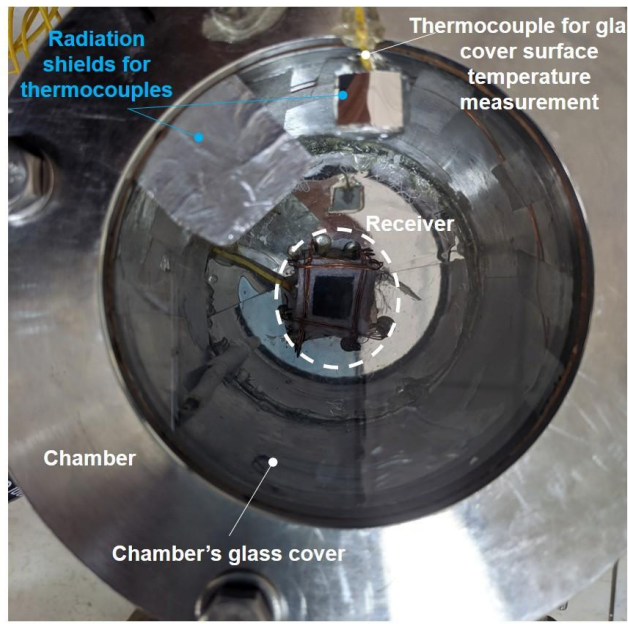

(a)

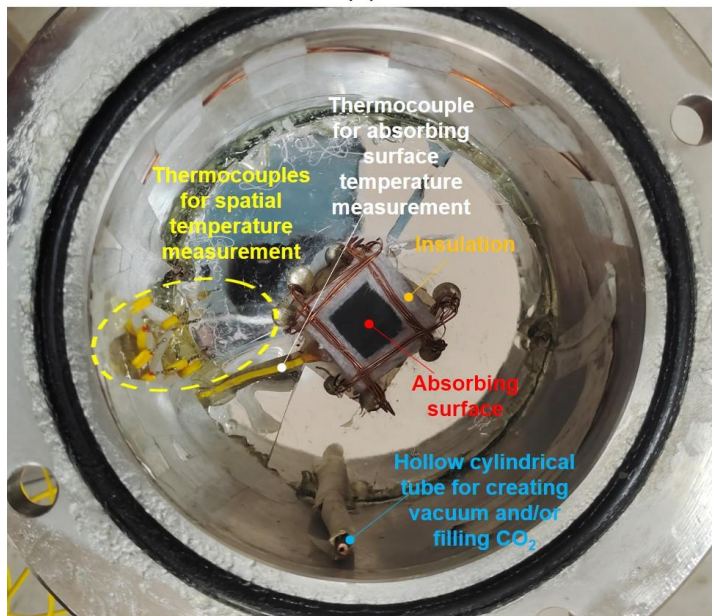

(b)

Fig. 11 Photograph showing (a) chamber with glass cover, and (b) chamber with glass cover removed.

### 3.2 Receiver design configurations

A receiver essentially consists of an absorber surface (which in the present work is either a black surface or a solar selective surface) surrounded by one or more OTTLMs (vacuum, air, $CO_2$, THM, aerogel).

The absorber surface is sandwiched between two insulating layers, where the bottom insulation is totally opaque, and the top insulation is also opaque (Pyrogel® XTE, Aspen Aerogels Inc.) but with an optically transparent window in the middle, so that the incoming irradiation is able to reach the absorber surface. For a given type of absorber surface, the aforementioned part of the construction of the receiver is common for all the investigated receiver designs (see Fig. 12). All the receiver configurations (with a given type of absorber surface) then differ in terms of the solid OTTLM (such as THM/aerogel) that masks the aforementioned transparent window and or the nature of the surrounding media (gaseous OTTLM such as air/vacuum/$CO_2$) within the chamber. Additionally, the orientation of the chamber (and hence that of the receiver) has also been varied to control the buoyancy driven bulk fluid motion (which in turn is dictated by spatial temperature gradients) of air/$CO_2$ within the chamber.

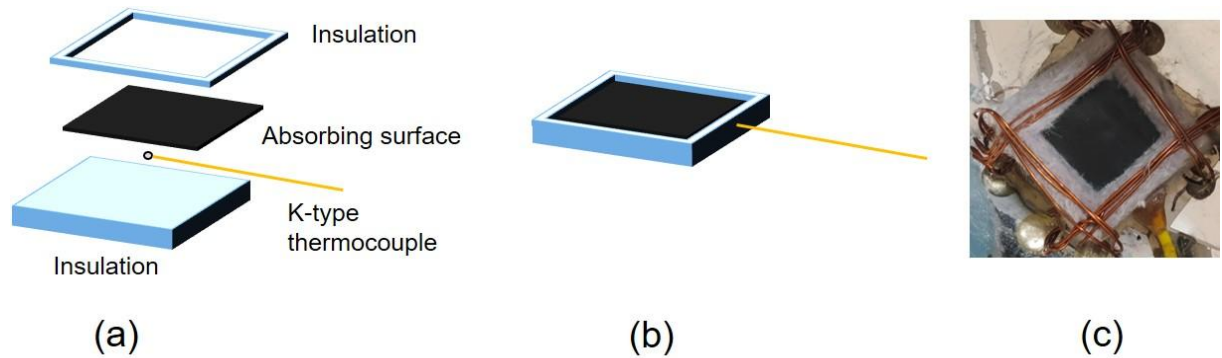

Fig. 12 Receiver construction part common to all investigated receiver designs. Schematic showing (a) the exploded view, (b) the assembled view, and (c) the actual photograph. It may also be noted that to ensure proper physical contact of the thermocouple with the absorber surface, copper winding has been employed to provide the necessary compressive force.

*3.2.1 Receiver designs with THMs/aerogel as the solid OTTLM*
In addition to the receiver design without any solid OTTLM (see Fig. 12(c)), two broad categories of receiver designs have been investigated with THM and aerogel as the OTTLM respectively.
In case of receiver design with THM as the solid OTTLM, two variants have been investigated. In one case, the coated side of the THM (ITO coated glass) is facing the incoming irradiation (see Fig. 13 (a)-(c), referred to as "$ITO_{IF}$") and in the second case, the coated side of the THM is facing the absorber surface of the receiver (see Fig. 13 (d)-(f), referred to as "$ITO_{RF}$").
Similarly, in case of receiver designs with aerogel as the solid OTTLM, two variants viz., one without (see Fig. 14 (a)-(c), referred to as "AE") and the other with reflecting ring (see Fig. 14 (d)-(f), referred to as "$AE_R$") have been investigated.

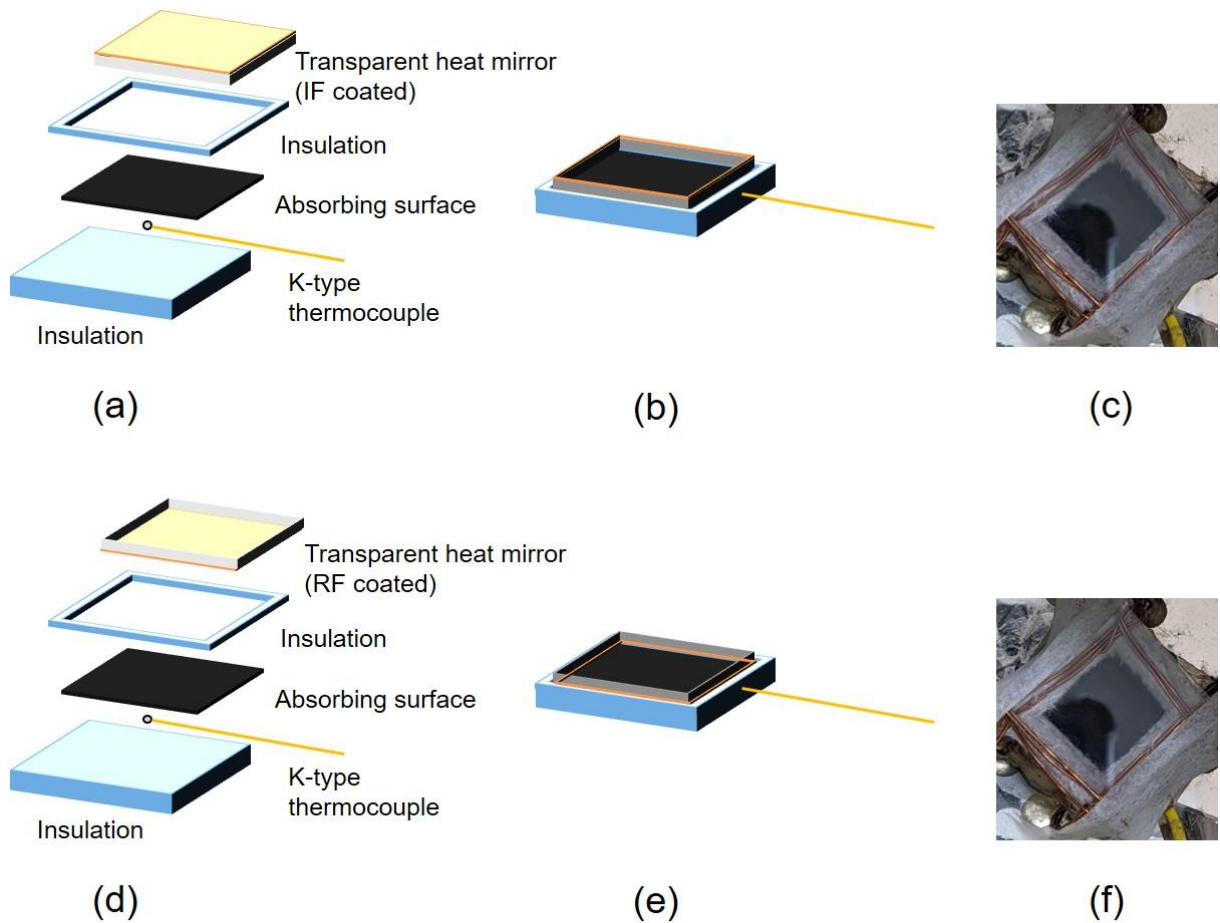

Fig. 13 Receiver design with THM (ITO coated glass) as the OTTLM. Coated side facing the incoming irradiation: Schematic showing (a) the exploded view, (b) the assembled view, and (c) the actual photograph. Coated side facing the absorber surface of the receiver: Schematic showing (d) the exploded view, (e) the assembled view, and (f) the actual photograph.

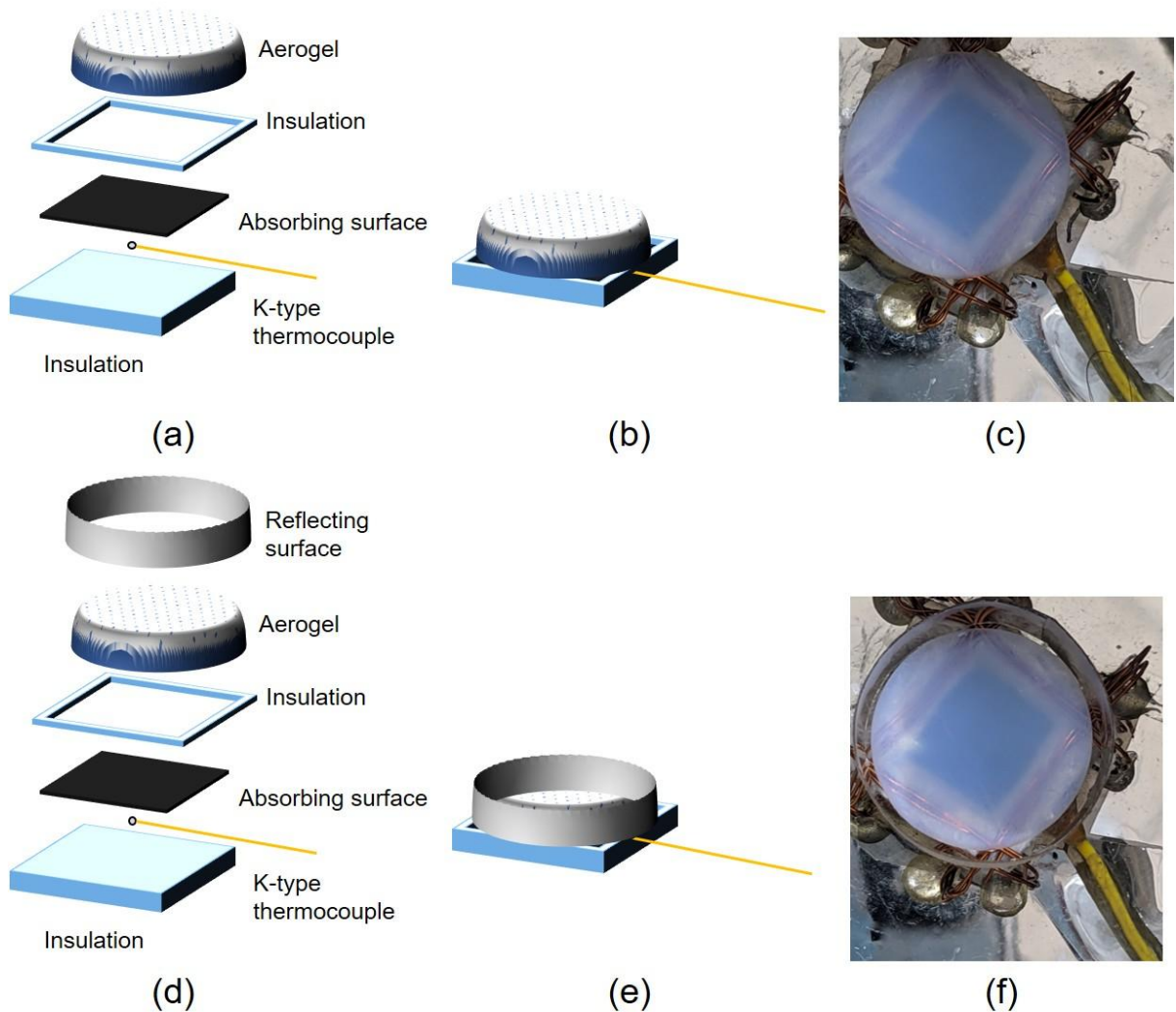

Fig. 14 Receiver design with aerogel (hydrophobic silica aerogel) as the OTTLM. Aerogel without reflecting ring: Schematic showing (a) the exploded view, (b) the assembled view, and (c) the actual photograph. Aerogel with reflecting ring (made from 3M$^{TM}$ solar mirror film): Schematic showing (e) the exploded view, (e) the assembled view, and (f) the actual photograph.

### *3.2.2 Controlling the bulk fluid motion within the chamber*
In addition to varying the absorber surface, solid OTTLMs, the receiver design configurations differ from each other in terms of the gaseous OTTLM surrounding the receiver (i.e., whether air/vacuum/$CO_2$). Furthermore, the magnitude of the bulk fluid motion within the chamber could be engineered by varying the orientation of the chamber and hence the receiver.

In particular, two extreme scenarios viz., irradiation from the top (see Fig. 15) and irradiation from the bottom (see Fig. 16) have been dealt with in the present work. The former orientation allows for significant bulk fluid motion due to unstable spatial temperature distribution. Herein, the absorber surface being at the bottom (and irradiation being from the top) favours initiation of buoyancy driven bulk fluid motion. On the other hand, in the latter orientation, the absorber is at the top (and irradiation is from the bottom) which results in temperature inversion that inhibits bulk fluid motion. Furthermore, it may be noted that for bottom irradiation experiments, a stand has been made such that the distance between the absorber surface and the solar simulator opening aperture remains the same for both top and bottom irradiation experiments.

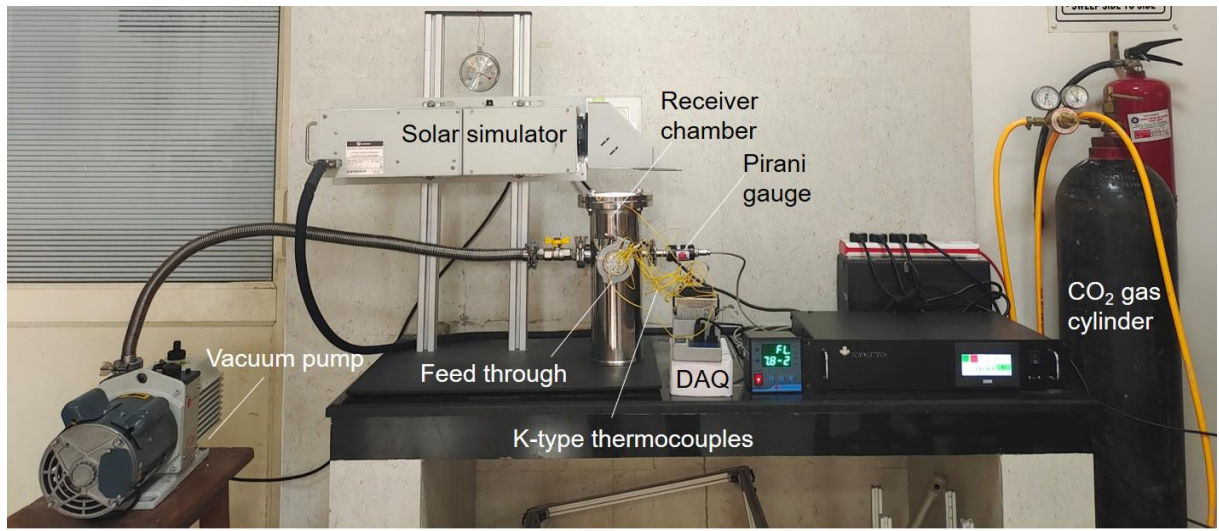

(a)

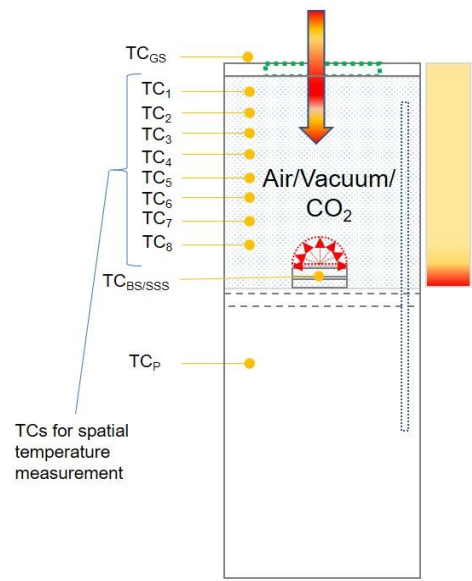

(b)

Fig. 15 Irradiation from the top. (a) Photograph of the entire experimental setup, and (b) schematic diagram of the receiver chamber.

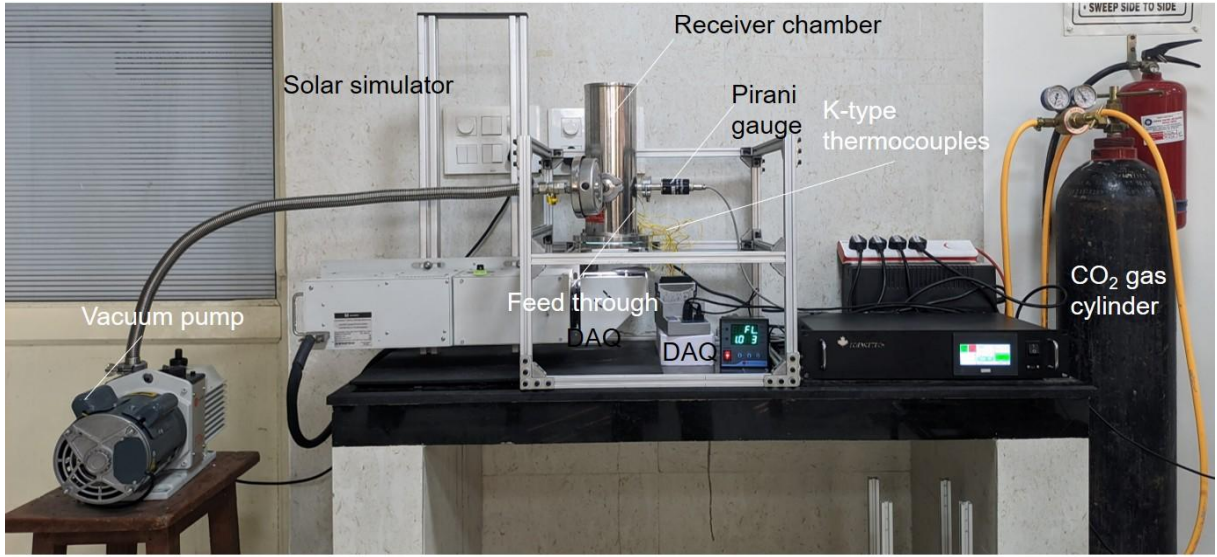

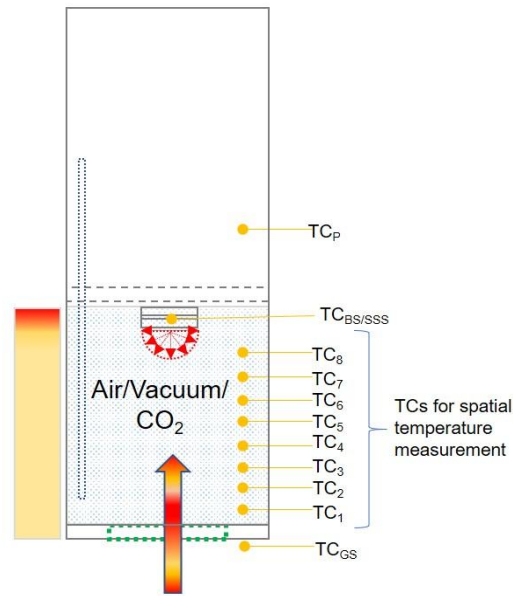

Fig. 16 Irradiation from the bottom. (a) Photograph of the entire experimental setup, and (b) schematic diagram of the receiver chamber.

**3.3 Scientific rationale for conducting optical heating experiments**
To better understand the operative heat transport mechanisms and to investigate (and quantify) the efficacy of different absorber surfaces and OTTLMs, it is imperative to single out the two key attributes viz., absorption capability and thermal losses.

To this end, optical heating experiments have been so designed such that a given receiver design configuration is optically irradiated until steady state is reached; subsequently it is allowed to cool to ambient temperature. The aforementioned processes allow determination of the absorber surface, glass cover and spatial temperatures within the chamber under steady state conditions (i.e., stagnation temperatures). Furthermore, through cooling curves the magnitude of the heat transfer coefficient is ascertained.

### 3.3.1 Determining stagnation temperatures through heating curves

Optical heating experiments have been carried out until near steady state condition is reached. In the present work, optical heating has been carried out until the rate of increase in absorber surface temperature reduced to approximately 0.1°C/min and 0.3°C/min for receiver design configurations involving air/$CO_2$ and vacuum respectively.

*Relevance of absorber surface stagnation temperature*

The magnitude of the absorber surface stagnation temperature (in particular) gives a fair idea about both the absorption capability of the absorber surface as well as quantum of thermal heat losses taking place from the absorber surface.

Absorption capability in turn is dictated by the optical characteristics of both the absorber surface as well as the OTTLM in the UV-Vis-NIR region of the electromagnetic spectra. On the other hand, thermal losses from the absorber surface depend on a multitude of factors. For instance, the magnitude of radiation losses depends on the emissivity of the absorber surface and optical characteristics of the OTTLM in the MIR of the electromagnetic spectra. Conduction losses depend on the conductivity of the OTTLM, whereas, the magnitude of the bulk fluid motion within the chamber dictates the magnitude of the convection losses.

*Relevance of spatial stagnation temperatures*

Nature of the temperature distribution in the space between the absorb surface and the glass cover gives an estimate of the magnitude of the bulk fluid motion within the space. This is particularly important to ascertain the extent of convection losses taking place from the absorber surface.

### 3.3.2 Determining heat loss coefficient through cooling curves

The rate at which the absorber surface cools down (determined through transient cooling curves) gives an estimate of the combined heat loss coefficient ($h_{comb}$, mathematically defined by Eq. 7). The magnitude of the aforementioned heat loss coefficient has the contribution from the heat loss via all the three heat transfer modes (namely conduction, convection, and radiation). However, through careful selection of the absorber surface, OTTLM, and control of the bulk fluid motion, it is possible to estimate the extent of contribution of each of the three modes of heat transfer.

$$h_{comb} = \left(\frac{1}{\tau_{surf}}\right) \times \left(\frac{m_{surf} \times c_{p,surf}}{A_{surf}}\right) \qquad (7)$$

where, $m_{surf}$, $c_{p,surf}$, $A_{surf}$ are the mass, specific heat, and surface area of the absorber surface; $\tau_{surf}$ is the heat transfer time constant (which in turn is defined by Eq. 8).

$$\frac{1}{\tau_{surf}} = \ln(\theta) \times \left(\frac{-1}{t}\right) \qquad (8)$$

where, $\theta = \frac{T_{amb} - T}{T_{amb} - T_{max}}$, and $t$ is the time elapsed. $T_{amb}$, $T$, and $T_{max}$ being ambient, instantaneous, and maximum absorber temperatures.

## 4. THERMAL PERFORMANCE OF ABSORBING MATERIALS AND OTTLMs UNDER OPTICAL HEATING

For a comprehensive and holistic comparison of the efficacy of various absorber surfaces and OTTLMs, receiver design configurations have been classified into four broad categories.

Wherein, the absorber surface and direction of irradiation are fixed for a given category of receiver design configuration, and efficacies of various OTTLMs are investigated.

**4.1 Black surface irradiated from the top**
Herein, the absorber surface is "spectrally black" and the direction of irradiation is from the top. This configuration represents the "worst case scenario", wherein, thermal losses via convection and radiation are expected to be at their maxima. This could be understood from the fact that emissivity of the "spectrally black" surface being close to unity results in emission to peak at a given absorber surface temperature. Furthermore, the absorber being at the bottom and irradiation being from the top, shall result in a temperature gradient which favours bulk fluid motion and hence convection losses from the absorber surface.
Figure 17 (a) shows the steady state absorber surfaces temperatures for the said configuration as a function of the OTTLM employed.
To aid comparison, the said broad category of the receiver design configuration has been divided into two sub-categories. One, in which, the chamber houses air/$CO_2$ and the other being the one in which vacuum is maintained in the chamber. Clearly, absorber stagnation temperatures are significantly higher in the case that involves vacuum in the chamber (see Fig. 17 (a)). This could be understood from the fact that in case of vacuum, there is negligibly small bulk fluid motion within the chamber and hence resulting in low convection losses from the absorber surface. In this regard, the fact that the spatial temperature distribution is nearly linear and the highest temperatures being in the vicinity of the absorber surface is a clear testimony that heat diffusion is indeed the key heat transport mechanism (in addition to the radiation) in case of vacuum in the chamber. On the other, in case of air/$CO_2$ in the chamber, highest temperatures occur away from the absorber surface – implying "buoyancy driven convective flows" within the chamber. Additionally, thermal conductivity being a strong function of the air pressure also plays a role in dictating the thermal losses via conduction. Thermal conductivity being relatively low in case of vacuum, results in lower conduction losses as well. Now that we have been able to decipher the heat transport mechanisms operational in case of receiver configurations involving air/$CO_2$ and vacuum in the chamber, let us now zoom into each of the aforementioned configurations in greater detail.

*4.1.1 Receiver design configurations with air/$CO_2$ in the chamber*
Receiver design configuration with air in the chamber and no solid OTTLM has been taken as the "base case" and all other receiver design configurations have been compared against this "base case". Figure of merit ($\delta T_{abs}(\%)$), mathematically defined by Eq. (9) gives the percentage enhancements in absorber stagnation temperatures relative to the "base case".

$$\delta T_{abs}(\%) = \frac{T_{abs} - T_{abs}|_{Air,A}}{T_{abs}|_{Air,A}} \times 100 \tag{9}$$

This, in essence, quantifies the efficacy of a given OTTLM relative to air.
Let us first consider the case of $CO_2$, herein, owing to the absorption bands in the infrared region of the electromagnetic spectrum, $CO_2$ helps in blocking some part of the radiations emitted from the absorber surface and hence helps in curbing radiation losses from the absorber. Although, the magnitude of enhancement ($\delta T_{abs}(\%)$) is small, but is of the same order of magnitude as for more sophisticated OTTLMs (such as ITO coated glass/aerogel). Moreover, looking carefully into the inset of Fig. 17(b) reveals that indeed $CO_2$ outperforms the configuration involving aerogel (without reflecting ring) i.e., the "AE,A" case. This may be attributed to the fact that scattering of the incident radiation by the aerogel – resulting in loss of optical energy reaching the absorber surface. However, circumscribing the aerogel with a reflecting ring (i.e., AE$_{R,A}$ case) reduces the optical loss due to scattering, (in the Vis-NIR region) resulting in increased incident radiation to reach the absorber surface and hence enhanced absorber surface temperatures.

Next, let us analyse THMs (ITO coated glass substrate) with air in the chamber. Herein, the thermal losses happen parallelly in the following two ways: thermal losses happening directly from the absorber surface owing to the bulk fluid motion (convection losses) and the thermal losses happening (primarily) via radiation, involving radiative exchange between the absorber and the ITO coated substrate which is finally lost from the other side of the ITO coated substrate surface both via radiation and convection. Herein, the absorber surface temperatures being not significantly high, renders thermal losses via the later heat transfer mode (i.e., radiation exchange between the absorber and the ITO coated surface) itself to be relatively low in magnitude. Hence, pointing out that the heat loss is primarily dictated by the convection mode of heat transfer (due to bulk fluid motion directly from the absorber surface).

When the coated side is facing the incident irradiation (i.e., "ITO$_{IF, A}$" case), the radiation exchange between the absorber surface and the ITO coated surface is larger than the case where the coated side is facing the absorber surface (i.e., ITO$_{RF,A}$ case). In other words, the latter case (i.e., "ITO$_{RF,A}$" case) offers more resistance for thermal loss to happen via radiation heat transfer mode. Hence, paving the way for heat loss to happen primarily via convection in ITO$_{RF,A}$ case. This points out that the ITO coated glass configuration that has lower convective losses shall have higher absorber surface stagnation temperature ("ITO$_{IF,A}$" configuration in the present case).

### *4.1.2 Receiver design configurations with vacuum in the chamber*

Now that we have assessed the efficacy of various solid OTTLMs with air/CO$_2$ in the chamber; next, we delve deeper into the subcategory wherein there is vacuum in the chamber. Firstly, the case wherein there is no solid OTTLM and with vacuum in the chamber is considered. Herein, the enhancements in stagnation temperature ($\delta T_{abs}(\%)$) on the order of 46% (relative to the "base case" i.e., Air$_A$) is achieved. This is attributed to significant reduction in bulk fluid motion at low vacuum pressures ($\sim 7 \times 10^{-2}$ mbar in the present work) and hence negligible small thermal losses via convection mode of heat transfer. Furthermore, magnitude of conduction losses is also small owing to reduction in thermal conductivity with decrease in pressure. Therefore, the thermal losses from the absorber surface predominantly occur via radiation mode of heat transfer (as vacuum is nearly transparent to the IR radiations emitted by the absorber surface).

Next, considering the case wherein ITO coated glass substrate is employed as the solid OTTLM and there is vacuum in the chamber. Here, enhancements are much higher ($\delta T_{abs}(\%) \approx 57\% - 58\%$, relative to the "base case" i.e., Air$_A$) than that observed in case when there is no solid OTTLM and there is vacuum in the chamber. This could be understood from the fact that ITO coated glass substrate interacts with the IR radiations emitted from the absorber surface, thereby blocking it (via absorption and reflection mechanisms) to escape.

In case of aerogel as the solid OTTLM and with vacuum in the chamber, higher enhancements in absorber surface temperatures are recorded ($\delta T_{abs}(\%) \approx 63\%$, relative to the "base case" i.e., Air$_A$). Herein, vacuum helps in curbing the convective losses and conduction and radiation losses from the absorber surface are contained owing to the low thermal conductivity and high optical thickness (in the IR region) of the aerogel. As noted earlier (subsection 4.1.1), the enhancements are further increased by circumscribing the aerogel with a reflective ring ($\delta T_{abs}(\%) \approx 66\%$, relative to the "base case" i.e., Air$_A$).

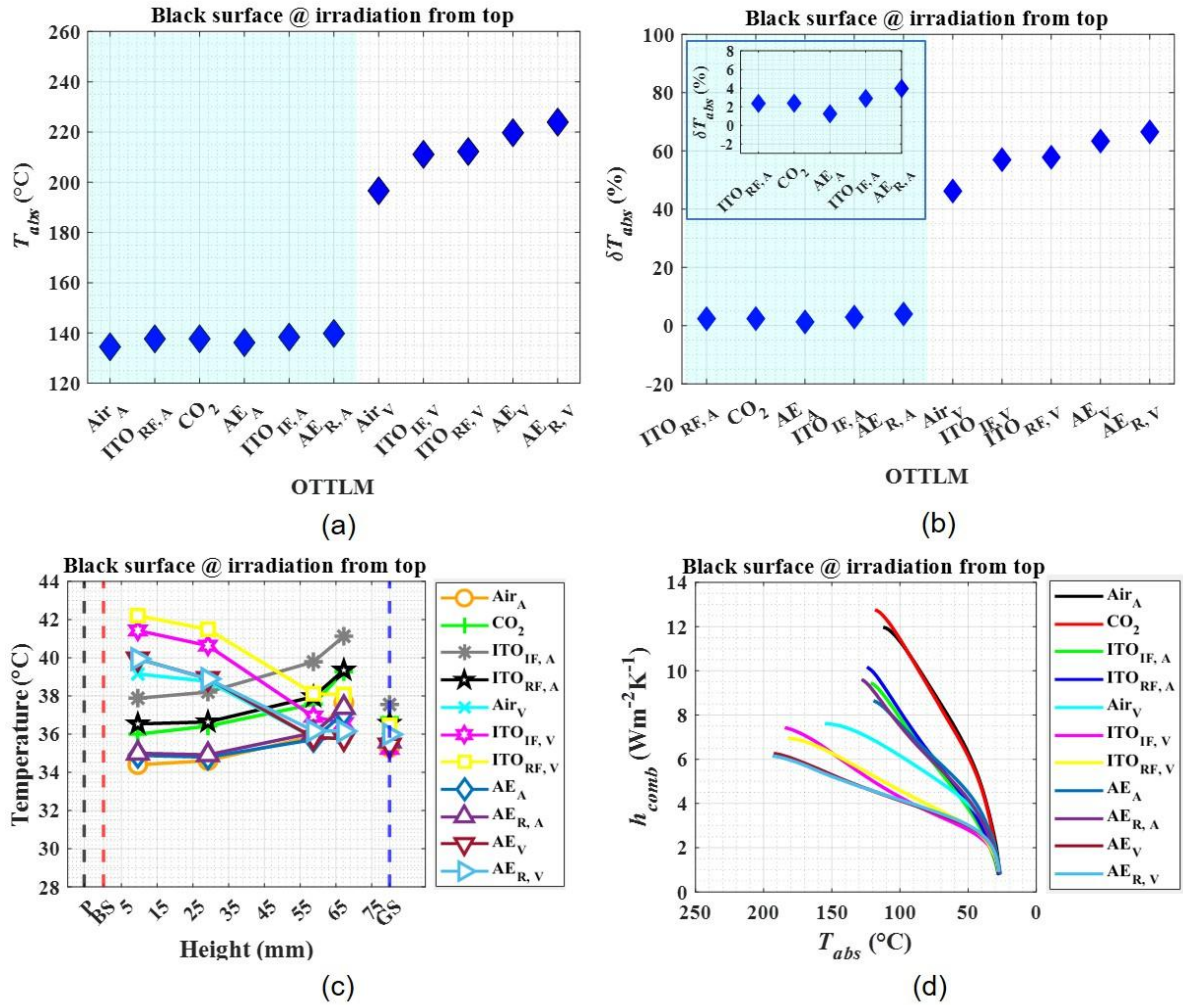

Fig. 17 Black surface irradiated from top: (a) Absorber surface steady state (stagnation) temperatures for various OTTLMs, (b) Enhancements in absorber surface stagnation temperatures relative to the "base case", (c) Spatial temperature distribution within the chamber, and (d) Combined heat transfer coefficient as a function of absorber surface temperature.

### 4.2 Solar selective surface irradiated from the top

This configuration is similar to the previous configuration (i.e., "Black surface irradiated from the top") except for the absorber surface material. As the said configuration employs "solar selective" surface instead of the "spectrally black" surface as the absorber surface, thermal losses via radiation are bound to be lower than that for the corresponding "spectrally black" surface case.

#### *4.2.1 Receiver design configurations with air/$CO_2$ in the chamber*

Herein, as the absorber surface is "solar selective", and the fact that temperatures are not significantly high, the thermal losses are primarily dictated by the magnitude of the bulk fluid motion and hence the convection mode of heat transfer. Furthermore, the optical transmittance of the intervening OTTLM (in the Vis-NIR region) affects the magnitude of the incident radiation that is able to reach the absorber surface.

Firstly, considering the case when there is $CO_2$ in the chamber. Herein, due to higher molecular weight of $CO_2$ (relative to air), bulk fluid motion is small compared that in case of air in the

chamber – therefore, resulting in lower convection losses from the absorber surface and higher absorber surface stagnation temperature ($\delta T_{abs}(\%) \approx 3.3\%$, relative to the "base case" i.e., $Air_A$). Furthermore, $CO_2$ has some absorption bands in the IR region of the electromagnetic spectra, which also results in reduction in radiation thermal losses (although in a relatively small manner, as the absorbing surface is solar selective and temperatures are not significantly high). Next, analysing the case of ITO coated substrates with air in the chamber. ITO coated substrate interacts with the IR radiations emitted from the absorber surface (which otherwise would have been able to escape in the case where there is no solid OTTLM i.e., in case of "base case" ($Air_A$)). However, as noted earlier, the magnitude of radiation losses being relatively small, the enhancements relative to the "base case" are negligibly small.

Although, like ITO coated substrate, aerogel does help in curbing thermal losses, however, the magnitude of these enhancements are very low (owing to the fact that the absorber temperature being low and the surface being solar selective). Furthermore, the fact that aerogel scatters the incoming radiation – this results in overall negative enhancements relative to the "base case".

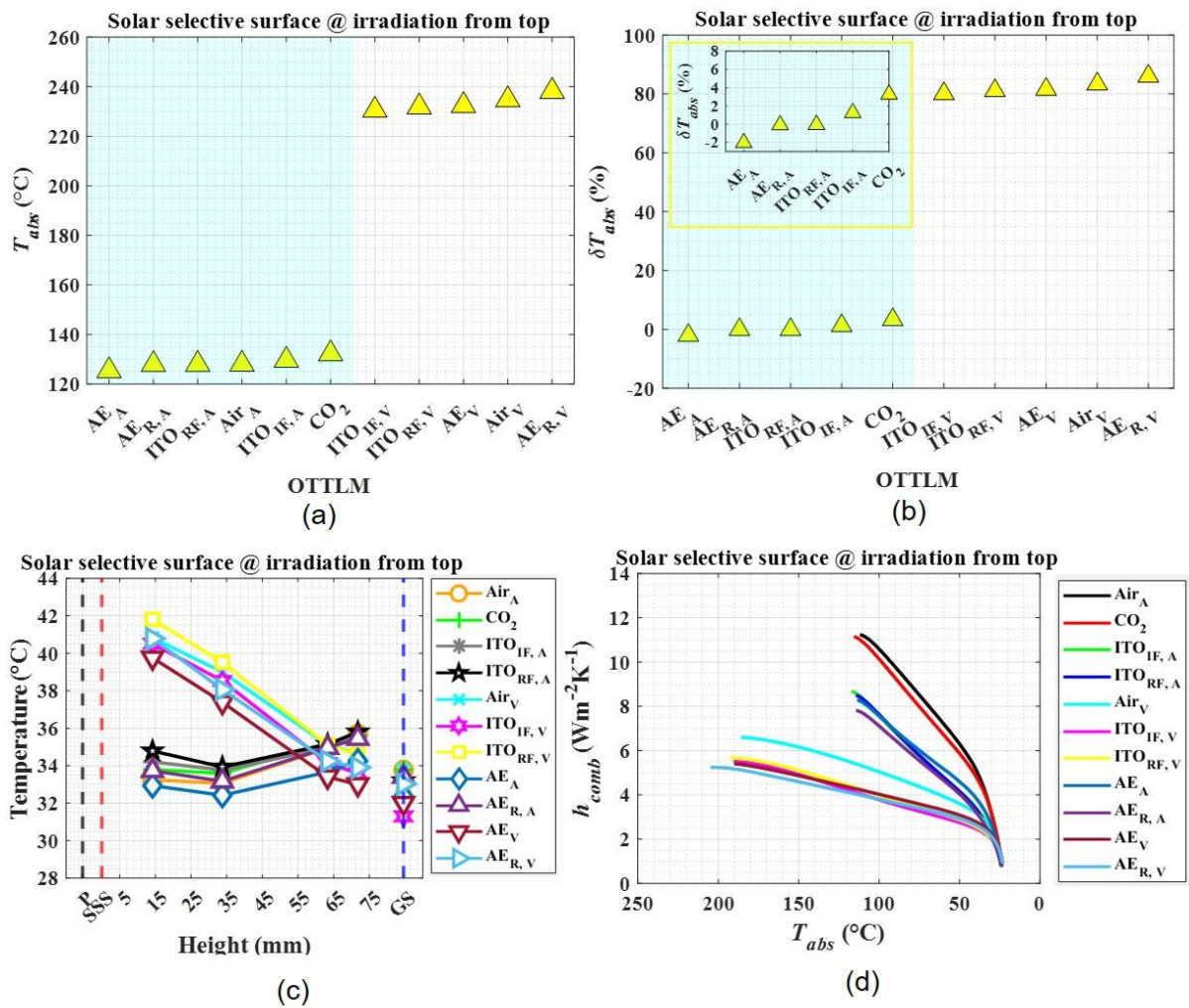

Fig. 18 Solar selective surface irradiated from top: (a) Absorber surface steady state (stagnation) temperatures for various OTTLMs, (b) Enhancements in absorber surface stagnation temperatures relative to the "base case", (c) Spatial temperature distribution within the chamber, and (d) Combined heat transfer coefficient as a function of absorber surface temperature.

*4.2.2 Receiver design configurations with vacuum in the chamber*
Next, considering the case of vacuum in the chamber, among all the receiver designs in this subcategory, the configuration involving aerogel with reflective ring as the solid OTTLM performs the best ($\delta T_{abs}(\%) \approx 86\%$, relative to the "base case" i.e., $Air_A$). This may be attributed to combined thermal loss control through vacuum (which curbs convection and conduction losses) and aerogel (which curbs radiation and conduction losses). The case involving no OTTLM performs the second best ($\delta T_{abs}(\%) \approx 83\%$, relative to the "base case" i.e., $Air_A$). The remaining receiver design configurations (namely $ITO_{IF,V}$, $ITO_{RF,V}$, and $AE_{,V}$) have nearly similar enhancements ($\delta T_{abs}(\%) \approx 80\% - 81\%$, relative to the "base case" i.e., $Air_A$). Clearly, the enhancements in case of vacuum in chamber with and without solid OTTTLMs are not significantly different from each other owing to the fact that solar selective surface itself ensures low radiation losses at a given surface temperature, therefore, effect of solid OTTLM is not displayed in a pronounced manner.

**4.3 Black surface irradiated from the bottom**
Herein, the direction of irradiation is from bottom and the absorbing surface is at the top, therefore, inhibiting bulk fluid motion (clearly apparent from the linear spatial temperature distribution within the chamber, see Fig. 19 (c)) and hence reducing convection losses from the absorber surface. However, as the absorbing surface being spectrally black, shall result in high radiation at a given absorber surface temperature.

*4.3.1 Receiver design configurations with air/$CO_2$ in the chamber*
Firstly, considering the case wherein, there is no solid OTTTM and the chamber contains $CO_2$. Herein, enhancements ($\delta T_{abs}$) on the order 6.3% (relative to the "base case" i.e., $Air_A$) are observed (see inset in Fig. 19(b)). This is significantly higher than that observed in the corresponding case of $CO_2$ with irradiation from the top (see inset in Fig. 17 (b)). The increased enhancement may be attributed to the significantly reduced bulk fluid motion, which ultimately results in reduction of convective losses from the absorber surface.

*4.3.2 Receiver design configurations with vacuum in the chamber*
Next, considering the cases wherein there is vacuum in the chamber. With no solid OTTLM, enhancements ($\delta T_{abs}$) on the order 50% (relative to the "base case" i.e., $Air_A$) are observed (see Fig. 19(b)). Masking with ITO coated glass substrate further enhances the enhancements ($\delta T_{abs}$) to approximately $59\% - 61\%$ (relative to the "base case" i.e., $Air_A$). Although trends similar to that observed for the corresponding cases of heating from top are observed, however, the magnitude of the enhancements are higher in the present cases owing to reduced bulk fluid motion and hence reduced convection losses.

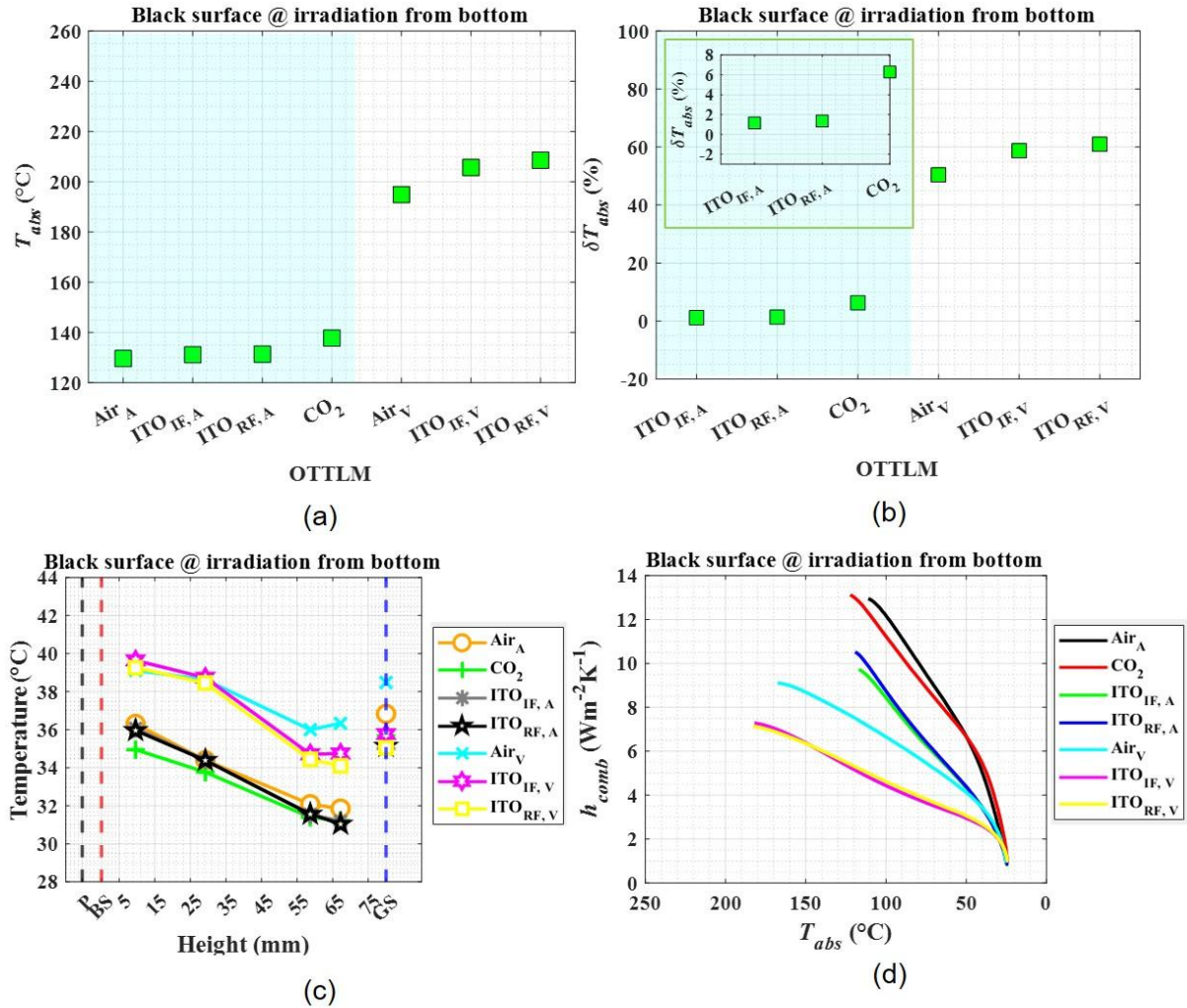

Fig. 19 Black surface irradiated from bottom: (a) Absorber surface steady state (stagnation) temperatures for various OTTLMs, (b) Enhancements in absorber surface stagnation temperatures relative to the "base case", (c) Spatial temperature distribution within the chamber, and (d) Combined heat transfer coefficient as a function of absorber surface temperature.

### 4.4 Solar selective surface irradiated from the bottom

As the irradiation is from the bottom (which ensures minimal bulk fluid motion, clearly apparent from the linear spatial temperature distribution within the chamber, see Fig. 20 (c)) and the absorbing surface is solar selective (which ensures minimal radiative losses at a given absorber surface temperature), the thermal losses are bound to be lowest among all the considered receiver design configuration.

#### 4.4.1 Receiver design configurations with air/CO$_2$ in the chamber

With $CO_2$ in the chamber and the absorber surface being solar selective, there is improvement in radiation thermal loss mitigation. Further, as the irradiation is from the bottom, the bulk fluid motion is also inhibited. The aforementioned conditions result in overall reduction of thermal losses and enhancements ($\delta T_{abs}$) on the order 7% (relative to the "base case" i.e., Air$_A$) are observed (being highest among all the investigated CO$_2$ based receiver design configurations).

#### 4.4.2 Receiver design configurations with vacuum in the chamber

As the absorber surface is "solar selective", irradiation is from the bottom and the fact that there is vacuum in the chamber – all these have a synergistic effect of reducing the overall thermal losses from the absorber surface. Therefore, this receiver design configuration without any solid OTTTM results in highest enhancements (($\delta T_{abs} \approx 84\%$), relative to the "base case" i.e., Air$_A$) among all the corresponding receiver designs investigated in the present work. For receiver designs with solid OTTLMs, the optical transparency of the solid OTTLM in the Vis-NIR wavelength band and its reflectivity in the MIR shall dictate the magnitude of enhancements achieved. In case of ITO coated substrate kept on the absorber surface (with coating side facing the absorber surface i.e., ITO$_{RF,V}$), the optical loss in the Vis-NIR wavelength band is compensated by the thermal loss reduction in the MIR region. Thus, similar order of enhancements (($\delta T_{abs} \approx 84\%$), relative to the "base case" i.e., Air$_A$) as in the case with no solid OTTLM is observed. However, in case of ITO coated substrate with the coating side facing the incoming irradiation (i.e., ITO$_{IF,V}$), the effective IR emissivity being higher results in lower enhancements (($\delta T_{abs} \approx 79\%$), relative to the "base case" i.e., Air$_A$).

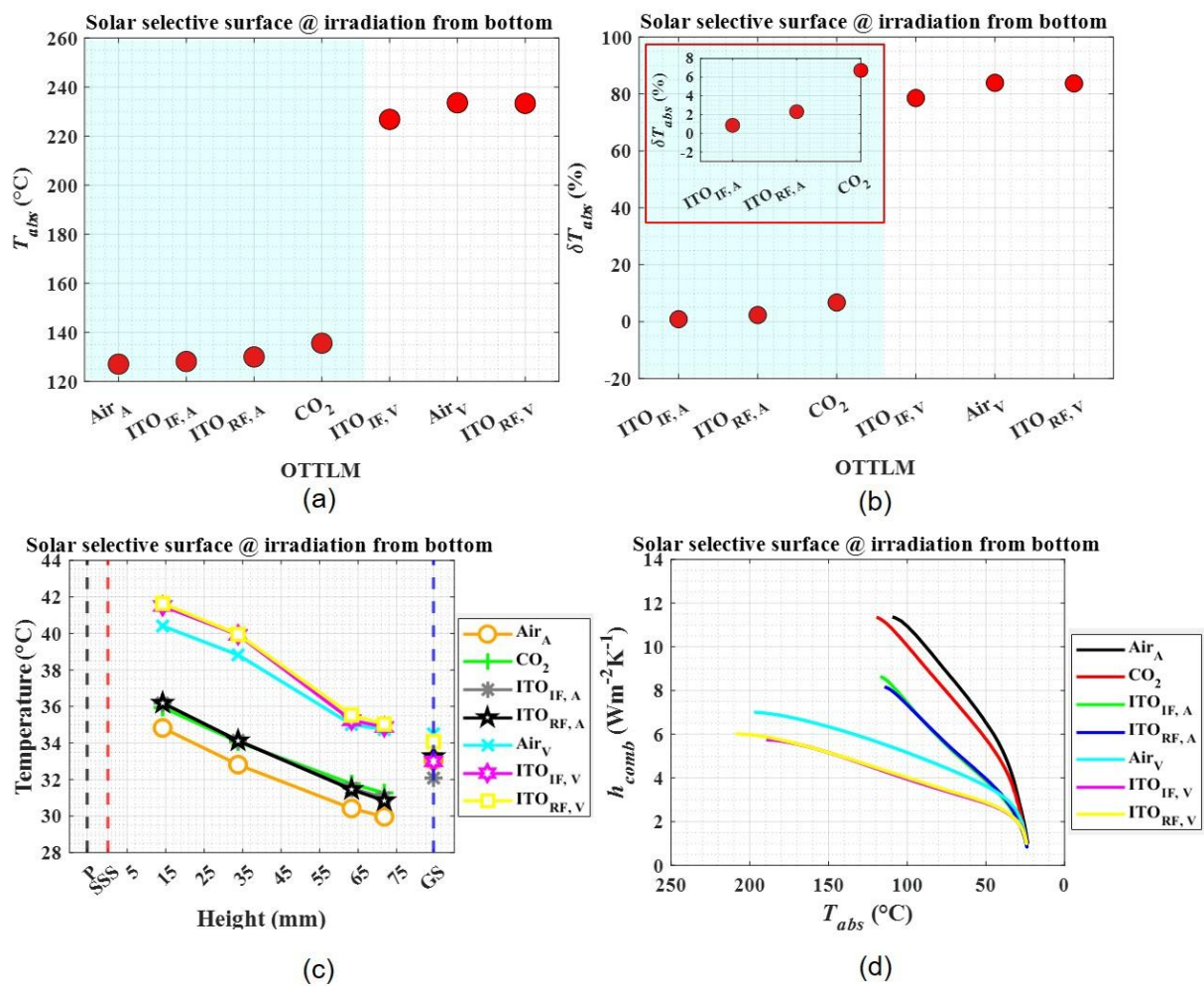

Fig. 20 Solar selective surface irradiated from bottom: (a) Absorber surface steady state (stagnation) temperatures for various OTTLMs, (b) Enhancements in absorber surface stagnation temperatures relative to the "base case", (c) Spatial temperature distribution within the chamber, and (d) Combined heat transfer coefficient as a function of absorber surface temperature.

# CONCLUSIONS AND FUTURE DIRECTIONS

Through present work a comprehensive and holistic attempt has been made to identify, understand and quantify the modes of heat transfer operational in a host of OTTLMs (Air, vacuum, THMs, aerogel, $CO_2$). Detailed and careful experimental modelling reveals that each investigated OTTLM has unique opto-thermal characteristics. For enhancing the thermal mitigation capability, it is imperative to carefully choose combinations of these OTTLMs so that there is a synergy between their thermal loss mitigation attributes. Furthermore, location of the absorber surface (relative to the direction of irradiation) dictates the magnitude of the bulk fluid motion within the chamber and hence the convective losses from the absorber surface. Among all the OTTLMs investigated in the present work and under the given operating conditions, aerogel possesses almost all the desired opto-thermal characteristics required for solar thermal applications. However, evacuation of air is necessary to realize its true thermal loss mitigation potential.

Although potentially promising candidates, sophisticated OTTLMs such as aerogels and THMs have some limitations (in relation to their synthesis, scalability and cost) which need to be addressed to realize their deployment in real world systems. Therefore, there exists a pressing need to explore other combinations of OTTLMs as well to realize simpler and inexpensive alternatives.

# ACKNOWLEDGEMENTS

This work is supported by DST-SERB (under Sanction order no. CRG/ 2021/003272). Authors also wish to acknowledge the support provided by Micro and Nano Characterization Facility (MNCF) at CeNSE, Indian Institute of Science (under INUP), Bengaluru; Institute Instrumentation Centre (IIC) at Indian Institute of Technology Roorkee; Mechanical Engineering Department and School of Physics and Material Science at Thapar Institute of Engineering & Technology, Patiala.

# NOMENCLATURE

English Symbols:

| | |
|---|---|
| $c_p$ | specific heat [J kg$^{-1}$ K$^{-1}$] |
| $E_{b,\lambda}$ | spectral black body emissive power [Wm$^{-2}$μm$^{-1}$] |
| $h$ | heat transfer coefficient [Wm$^{-2}$K$^{-1}$] |
| $K_{abs}$ | absorption coefficient [m$^{-1}$] |
| $L$ | physical pathlength |
| $m$ | mass [kg] |
| $P$ | pressure [Nm$^{-2}$] |
| $R$ | universal gas constant [Jmol$^{-1}$K$^{-1}$] |
| $S_\lambda$ | spectral irradiation [Wm$^{-2}$μm$^{-1}$] |
| $H$ | height of channel [m] |
| $T$ | temperature [K] |
| $t$ | time [s] |
| $w$ | mass pathlength [kgm$^{-2}$] |
| $x$ | mole fraction |

Greek Symbols:

| | |
|---|---|
| $\alpha$ | absorptivity |
| $\varepsilon$ | emissivity |
| $\lambda$ | wavelength |
| $\theta$ | dimensionless temperature |
| $\tau$ | time constant |

Subscript:
*A*　　air
*AE*　　aerogel
*amb*　　ambient
*b*　　black body
*eff*　　effective
*IF*　　irradiation facing
*max*　　maximum
*R*　　reflective
*RF*　　receiver facing
sw　　solar-weighted
*V*　　vacuum
*λ*　　spectral

Abbreviations and acronyms:
OTTLM　　optically transparent thermal loss mitigator
IR　　infrared
ITO　　indium tin oxide
NIR　　near infrared   Black surface-heat mirror receiver design
MIR　　mid infrared
THM　　transparent heat mirror
UMA　　universal measurement accessory
Vis　　visible

**REFERENCES**

[1] Dudley, Vernon E., Gregory J. Kolb, A. Roderick Mahoney, Thomas R. Mancini, Chauncey W. Matthews, M. I. C. H. A. E. L. Sloan, and David Kearney. *Test results: SEGS LS-2 solar collector*. No. SAND94-1884. Sandia National Lab.(SNL-NM), Albuquerque, NM (United States), 1994.
[2] McEnaney, Kenneth, Lee Weinstein, Daniel Kraemer, Hadi Ghasemi, and Gang Chen. "Aerogel-based solar thermal receivers." *Nano Energy* 40 (2017): 180-186.
[3] Günay, A. Alperen, Hannah Kim, Naveen Nagarajan, Mateusz Lopez, Rajath Kantharaj, Albraa Alsaati, Amy Marconnet, Andrej Lenert, and Nenad Miljkovic. "Optically transparent thermally insulating silica aerogels for solar thermal insulation." *ACS applied materials & interfaces* 10, no. 15 (2018): 12603-12611.
[4] Zhao, Lin, Bikram Bhatia, Sungwoo Yang, Elise Strobach, Lee A. Weinstein, Thomas A. Cooper, Gang Chen, and Evelyn N. Wang. "Harnessing heat beyond 200 C from unconcentrated sunlight with nonevacuated transparent aerogels." *ACS nano* 13, no. 7 (2019): 7508-7516.
[5] Berquist, Zachary J., Kevin K. Turaczy, and Andrej Lenert. "Plasmon-enhanced greenhouse selectivity for high-temperature solar thermal energy conversion." *ACS nano* 14, no. 10 (2020): 12605-12613.
[6] Berquist, Zachary J., Andrew J. Gayle, Andrés Miranda Mañón, Victor Vogt, Keyi Kang Yao, Vishnu Ramasawmy, Kyle Wilke, Neil P. Dasgupta, and Andrej Lenert. "Large area transparent refractory aerogels with high solar thermal performance." *Solar Energy* 292 (2025): 113437.
[7] Fan, John CC, and Frank J. Bachner. "Transparent heat mirrors for solar-energy applications." *Applied optics* 15, no. 4 (1976): 1012-1017.

[8] Khullar, Vikrant, Prashant Mahendra, and Madhup Mittal. "Applicability of heat mirrors in reducing thermal losses in concentrating solar collectors." *Journal of Thermal Science and Engineering Applications* 10, no. 6 (2018): 061004.

[9] Khullar, Vikrant, Himanshu Tyagi, Todd P. Otanicar, Yasitha L. Hewakuruppu, and Robert A. Taylor. "Solar selective volumetric receivers for harnessing solar thermal energy." *Journal of Heat Transfer* 140, no. 6 (2018): 062702.

[10] Singh, Nirmal, and Vikrant Khullar. "Experimental and theoretical investigation into effectiveness of ZnO based transparent heat mirror covers in mitigating thermal losses in volumetric absorption based solar thermal systems." *Solar Energy* 253 (2023): 439-452.

[11] https://www.edmundoptics.in/f/acktar-light-absorbent-foil/14802/ (accessed on 6 April 2025)

[12] Edwards, D. K. "Absorption by infrared bands of carbon dioxide gas at elevated pressures and temperatures." *Journal of the Optical Society of America* 50, no. 6 (1960): 617-626.